\documentclass[english,a4paper,pre,aps,reprint]{revtex4-2}
\usepackage[T1]{fontenc}
\usepackage[latin9]{inputenc}
\usepackage{amsmath}
\usepackage{amssymb}
\usepackage{graphicx}

\makeatletter
\usepackage{natbib}

\makeatother

\usepackage{babel}
\begin{document}
\title{Stochastic entropy production associated with quantum measurement
in a framework of Markovian quantum state diffusion}
\author{Claudia L. Clarke and Ian J. Ford }
\affiliation{Department of Physics and Astronomy and London Centre for Nanotechnology,
University College London, Gower Street, London WC1E 6BT, U.K.}
\begin{abstract}
The reduced density matrix that characterises the state of an open
quantum system is a projection from the full density matrix of the
quantum system and its environment, and there are many full density
matrices consistent with a given reduced version. Without a specification
of relevant details of the environment, the time evolution of a reduced
density matrix is therefore typically unpredictable, even if the dynamics
of the full density matrix are deterministic. With this in mind, we
investigate a two level open quantum system using a framework of quantum
state diffusion. We consider the pseudorandom evolution of its reduced
density matrix when subjected to an environment-driven process that
performs a continuous quantum measurement of a system observable,
invoking dynamics that asymptotically send the system to one of the
relevant eigenstates. The unpredictability is characterised by a stochastic
entropy production, the average of which corresponds to an increase
in the subjective uncertainty of the quantum state adopted by the
system and environment, given the underspecified dynamics. This differs
from a change in von Neumann entropy, and can continue indefinitely
as the system is guided towards an eigenstate. As one would expect,
the simultaneous measurement of two non-commuting observables within
the same framework does not send the system to an eigenstate. Instead,
the probability density function describing the reduced density matrix
of the system becomes stationary over a continuum of pure states,
a situation characterised by zero further stochastic entropy production.
Transitions between such stationary states, brought about by changes
in the relative strengths of the two measurement processes, give rise
to finite positive mean stochastic entropy production. The framework
investigated can offer useful perspectives on both the dynamics and
irreversible thermodynamics of measurement in quantum systems. 
\end{abstract}
\maketitle

\section{Introduction}

In classical statistical mechanics, entropy quantifies  uncertainty
in the adopted configuration of a system when only partial detail
is available concerning the coordinates of the component particles.
This is a subjective uncertainty, a reflection of the personal state
of ignorance of a given observer. The capacity of an observer to predict
future behaviour when such a system is coupled to a similarly underspecified
environment is limited and their knowledge of the state worsens with
time, even if the dynamics are entirely deterministic. The total entropy
of the system and environment increases as a consequence. In many
situations such evolution can be associated with the dissipation of
potential energy into heat, and this underpins the role played by
entropy in the (19th century) second law of thermodynamics \citep{lebowitz1993boltzmann,albert2009time,Ford-book2013}.

The 21st century concept of entropy production, however, is based
on mechanics; specifically a consideration of the probabilities of
forward and backward sequences of events governed by an effective
stochastic dynamics. In this framework of `stochastic thermodynamics',
entropy change is the expectation value of a `stochastic entropy production',
clarifying a number of long standing conceptual issues \citep{seifert2008stochastic,harris2007fluctuation,SpinneyFord12a,spinney2012entropy,ford2015stochastic}.

\begin{figure}
\begin{centering}
\includegraphics[width=1\columnwidth]{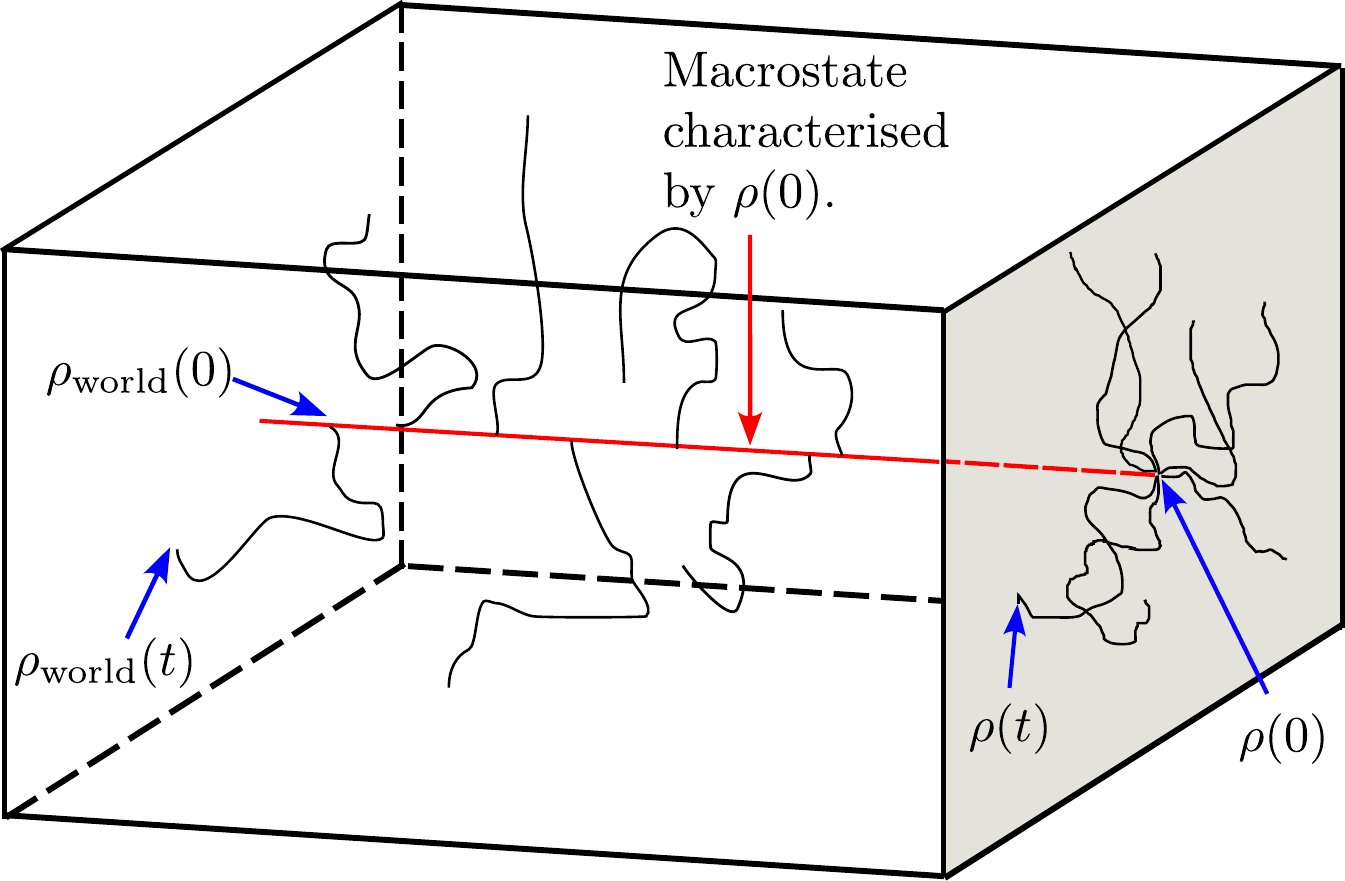}
\par\end{centering}
\caption{The box and the grey area represent the phase spaces of the density
matrix of the world $\rho_{{\rm world}}$ and of the reduced density
matrix $\rho$ of a constituent open system, respectively. Deterministic
trajectories $\rho_{{\rm world}}(t)$ that start at $t=0$ from a
macrostate subspace (shown as a red line) characterised by a given
initial value $\rho(0)$ of the reduced density matrix, can be manifested
as pseudorandom trajectories for $\rho(t)$ in the reduced phase space,
shown as projections onto the right hand face. \label{fig:Projection.}}
\end{figure}

The central aim of this paper is to employ entropy as a description
of uncertainty of the adopted configuration at the level of a reduced
density matrix in quantum mechanics. Putting aside the issue of quantum
measurement for the moment, the full density matrix of a system together
with its environment (a closed `world') evolves deterministically
according to the unitary dynamics of the von Neumann equation. This
can give rise to an evolution of the reduced density matrix describing
the system that preserves unit trace and positivity but allows changes
to von Neumann entropy and purity, corresponding to thermalisation
for example \citep{wiseman1996,breuer2007,weiss2012quantum}. But
the trajectory followed will be unpredictable if the complete initial
state of the world is not specified. It is natural to regard this
as producing an effective Brownian motion of the reduced density matrix,
and to concern ourselves with the associated entropy increase. The
concept is illustrated in Fig. \ref{fig:Projection.}. This intrinsic
unpredictability holds whether or not we introduce ideas of randomness
associated with quantum mechanical measurement.

In developing this idea, we view the reduced density matrix as an
analogue of classical system coordinates and hence as a physical description
of the quantum state, not merely as a vehicle for specifying probabilities
of projective measurements or a representation of a state of knowledge.
 But coordinates that describe the physical state of a system ought
not to change discontinuously, which would seem to raise difficulties
in connection with the instantaneous projections normally considered
to arise from quantum measurement. If the density matrix represents
a physical state we are therefore obliged to model quantum measurement
in a fashion that avoids discontinuous jumps. 

To this end, we pursue the idea that quantum measurement, namely the
adoption by a system of an eigenstate of an observable when interrogated
by a measuring device, is brought about by the deterministic dynamics
of the density matrix describing the system and its environment, of
which the measuring device would be a part. We explore the idea that
quantum measurement arises from the unitary dynamics of the world,
its apparent stochasticity being a consequence of a failure to specify
the initial degrees of freedom of the environment, or more precisely
those of a measuring device. Such an origin of stochasticity is reminiscent
of ideas employed in classical statistical mechanics.

In such a scheme, the evolution of the system under the influence
of its environment would be governed by a nonlinear dynamics with
attractors corresponding to the appropriate eigenstates. It is not
our aim here to derive such nonlinear dynamics from an underlying
evolution of the world. Instead, we seek a model of system dynamics
that has the desired effect, namely that the reduced density matrix
under measurement should evolve along continuous trajectories, terminating
at eigenstates. 

The modelling of `weak measurement' in quantum mechanics produces
continuous stochastic quantum trajectories \citep{brun2000continuous,jacobs2014quantum,jacobs2006straightforward}.
Random incremental changes in the state of an open system are brought
about by projective measurements of remote parts of the environment.
We shall employ this mathematical framework (without the associated
narrative of remote projective measurement) to represent the envisaged
nonlinear dynamical interactions between the system and environment
that guide the system towards eigenstates of observables under certain
conditions \citep{srednicki1994chaos}.

The framework known as quantum state diffusion (QSD) \citep{percival1998,strunz1996linear,gisin1992quantum,gisin1993quantum,strunz1999open}
is a broad category of open quantum system dynamics that can represent
the phenomenology of weak measurement. A continuous, Markovian, stochastic
evolution of the reduced density matrix emerges. More elaborate schemes
are also possible, for example involving non-Markovian dynamics. The
approach can be used to model the continuous measurement of an open
system that is consistent with strong projective measurements as a
limiting behaviour and is compatible with the Born rule. Measurement
in QSD is a process driven by specific system-environment coupling
and takes place without discontinuities \citep{jordan2013,Vinjamampathy16,kammerlander2016coherence}.
This is a quantum dynamics that resembles classical dynamics, but
where the dynamical variables are the elements of a reduced density
matrix. It combines both aspects of quantum evolution: determinism
of the von Neumann equation together with stochasticity representing
measurement or more general environmental effects \citep{Walls24}. 

The idea that quantum jumps are not instantaneous but merely very
rapid is not an unusual one \citep{Minev19} and the non-locality
of quantum mechanical evolution remains intrinsic to the interpretation.
Nevertheless, such a viewpoint is not without its controversies \citep{holland2005,WiseGamb08,hiley2019,roch2014}.
In particular, a suggestion that the quantum state represents a physical
configuration of the world might appear to conflict with various positions
taken in the fundamental interpretation of quantum mechanics, for
example those where a physical state (`reality') is considered to
be induced by the projective measurement process. Moreover, the supposed
`hidden variables' carried by the system and the environment, ignorance
of which gives rise here to the effective stochastic evolution, might
seem to conflict with the breakage of Bell inequalities and other
similar statistical results \citep{Groeblacher07,norsen2017a}. Resolution
of this issue might involve a deeper consideration of the implications
of determinism \citep{Hossenfelder20}. Alternatively, one could simply
regard quantum state diffusion as merely a mathematical framework
for modelling continuous pseudorandom quantum evolution.

If the evolution of the reduced density matrix can be modelled in
a fashion that avoids discontinuities then the concept of stochastic
entropy production in quantum mechanics can be introduced in a straightforward
way \citep{deffner2011nonequilibrium,leggio2013entropy,horowitz2013entropy,elouard2017probing,elouard2017role,dressel2017arrow,monsel2018autonomous,manikandan2019fluctuation,Belenchia20,matos2022}.
Entropy production arising from evolution that includes quantum jumps
can also be considered, but this introduces difficulties that manifest
as infinities in the change in system Gibbs entropy \citep{elouard2017role}.
We believe that such problems ought to be avoided if possible.

When dynamical variables evolve according to Markovian stochastic
differential equations (SDEs), or It\^o processes \citep{gardiner2009stochastic},
it is possible to derive a related It\^o process for the stochastic
entropy production \citep{spinney2012entropy}. This allows us to
compute a stochastic entropy production associated with individual
Brownian trajectories taken by the reduced density matrix of a system.
For situations where the system is guided towards an eigenstate of
an observable, we can compute the stochastic entropy production characterising
a process of measurement. 

A positive expectation value of such a stochastic entropy production
represents increased subjective uncertainty in the quantum state of
the world. Growth in uncertainty is natural since we model the evolution
using stochastic methods starting from an incompletely specified initial
state. The state of the system can become \emph{less} uncertain, a
necessary aspect of the performance of measurement, but uncertainty
with regard to the state of the rest of the world will increase by
a greater amount, thereby allowing the second law of thermodynamics
to be satisfied. It should be noted that stochastic entropy production
here does not correspond to a change in von Neumann entropy, which
instead describes the uncertainty of outcome when a system is subjected
to projective measurement in a specific basis. We comment further
on this in Section \ref{subsec:Contrast-with-von}.

In Section \ref{sec:Measurement-of} we develop these ideas in the
context of the measurement of a single observable in a two level quantum
system starting in a mixed state \citep{Schmidt}. Mean stochastic
entropy production is found to be positive and without limit as the
system is guided, asymptotically in time, into one or other of the
two eigenstates. We go on in Section \ref{sec:Simultaneous-measurement-of}
to consider the simultaneous measurement of two non-commuting observables
and show how the stochastic entropy production is finite, a consequence
of the inability of the dynamics, in this situation, to guide the
system into a definite eigenstate of either observable. 

We interpret the results in Section \ref{sec:Interpretation} and
summarise our conclusions in Section \ref{sec:Conclusions}, suggesting
that dynamics based on quantum state diffusion, with an interpretation
of the reduced density matrix as a set of physical properties of a
state, together with the use of stochastic entropy production to monitor
the process of eigenstate selection, can provide some conceptual clarification
of the quantum measurement problem \citep{norsen2017a}.

\section{Measurement of $\sigma_{z}$\label{sec:Measurement-of}}

\subsection{Dynamics}

We consider a two level system described by a reduced density matrix
(hereafter, simply a density matrix $\rho$) defined in a basis of
eigenstates $\vert\pm1\rangle$ of the $\sigma_{z}$ operator. Pure
states denoting occupation of one of the two levels correspond to
$\rho_{\pm}^{e}=\vert\pm1\rangle\langle\pm1\vert$. Starting in the
mixed state $\rho=\frac{1}{2}\left(a_{+}\rho_{+}^{e}+a_{-}\rho_{-}^{e}\right)$,
where $a_{\pm}$ are real coefficients, we use a quantum state diffusion
approach to model the stochastic evolution of the system into one
or other of the levels in accordance with the Born rule. 

We consider a minimal scheme \citep{jacobs2014quantum} employing
a rule for stochastic transitions given by
\begin{equation}
\rho\to S^{\pm}(\rho)=\rho^{\prime\pm}=\frac{M_{\pm}\rho M_{\pm}^{\dagger}}{{\rm Tr}\left(M_{\pm}\rho M_{\pm}^{\dagger}\right)},\label{eq:moves}
\end{equation}
with
\begin{equation}
M_{\pm}=\frac{1}{\sqrt{2}}\left(\mathbb{I}-\frac{1}{2}c^{\dagger}cdt\pm c\sqrt{dt}\right),\label{Kraus}
\end{equation}
where $c=\alpha_{z}\sigma_{z}$, with real scalar parameter $\alpha_{z}$
designated as the strength of measurement. The $M_{\pm}$ are examples
of \emph{Kraus operators}, and the map in Eq. (\ref{eq:moves}) often
appears in descriptions of physical transformations of a density matrix.
The probabilities for the selection of one of the two possible outcomes
$\rho^{\prime\pm}$ after an infinitesimal timestep $dt$ are 
\begin{equation}
p_{\pm}(\rho)={\rm Tr}\left(M_{\pm}\rho M_{\pm}^{\dagger}\right)=\frac{1}{2}\left(1\pm C\sqrt{dt}\right),\label{probabilities}
\end{equation}
where $C={\rm Tr}\left(\rho\left(c+c^{\dagger}\right)\right)$. The
quantum map in Eq. (\ref{eq:moves}) preserves the trace of $\rho$.
Furthermore, since the Kraus operators in Eq. (\ref{Kraus}) differ
incrementally from (a multiple of) the identity, the positive definiteness
of $\rho$ is maintained \citep{Walls24}. The operator identity
$M_{+}^{\dagger}M_{+}+M_{-}^{\dagger}M_{-}=\mathbb{I}$ is also satisfied.
This scheme defines a stochastic dynamics representing the effect
of a device interrogating the occupation of levels of the system,
whereby the eigenstates of $\sigma_{z}$ are stationary, i.e. $p_{+}(\rho_{+}^{e})=p_{-}(\rho_{-}^{e})=1$,
$p_{-}(\rho_{+}^{e})=p_{+}(\rho_{-}^{e})=0$, and $S^{+}(\rho_{+}^{e})=\rho_{+}^{e}$,
$S^{-}(\rho_{-}^{e})=\rho_{-}^{e}$.

The two possible increments $d\rho^{\pm}=\rho^{\prime\pm}-\rho$ available
in the timestep $dt$ under the dynamics are
\begin{align}
d\rho^{\pm} & =\left(c\rho c^{\dagger}-\frac{1}{2}\rho c^{\dagger}c-\frac{1}{2}c^{\dagger}c\rho\right)dt-\left(\rho c^{\dagger}+c\rho-C\rho\right)Cdt\nonumber \\
 & \qquad\pm\left(\rho c^{\dagger}+c\rho-C\rho\right)\sqrt{dt},\label{increments}
\end{align}
and by evaluating the mean and variance of this increment in $\rho$
it may be shown that the evolution can also be represented by the
It\^o process
\begin{equation}
d\rho=\left(c\rho c^{\dagger}-\frac{1}{2}\rho c^{\dagger}c-\frac{1}{2}c^{\dagger}c\rho\right)dt+\left(\rho c^{\dagger}+c\rho-C\rho\right)dW,\label{Ito}
\end{equation}
where $dW$ is a Wiener increment with mean $\langle dW\rangle=0$
and variance $\langle dW^{2}\rangle=dt$, with the brackets representing
an average over the stochasticity. Note that terms of higher order
than linear in $dt$ will be neglected throughout. A continuous evolution
of the stochastic variable $\rho$ driven by the infinitesimal stochastic
variable $dW$ has emerged, analogous to a random walk or Brownian
motion. This is what is meant by quantum state diffusion.

A process of averaging then leads to the standard Lindblad equation
\citep{lindblad}:
\begin{equation}
\frac{d\bar{\rho}}{dt}=c\bar{\rho}c^{\dagger}-\frac{1}{2}\bar{\rho}c^{\dagger}c-\frac{1}{2}c^{\dagger}c\bar{\rho},\label{Lindblad}
\end{equation}
with $\bar{\rho}=\langle\rho\rangle$. Such a deterministic equation
describes the average dynamical behaviour of an ensemble of density
matrices. The actual trajectory followed by a system as it responds
to external interactions, however, is specified by the stochastic
Lindblad equation (\ref{Ito}) \citep{moodley2009stochastic,yan2016stochastic}.
The environment disturbs the system in a manner represented by one
of the transformations or moves given in Eq. (\ref{eq:moves}), selected
at random with probabilities (\ref{probabilities}) that arise from
the underspecification of the environmental state and hence of $\rho_{{\rm world}}$
in Fig. \ref{fig:Projection.}.

\begin{figure}
\begin{centering}
\includegraphics[width=1\columnwidth]{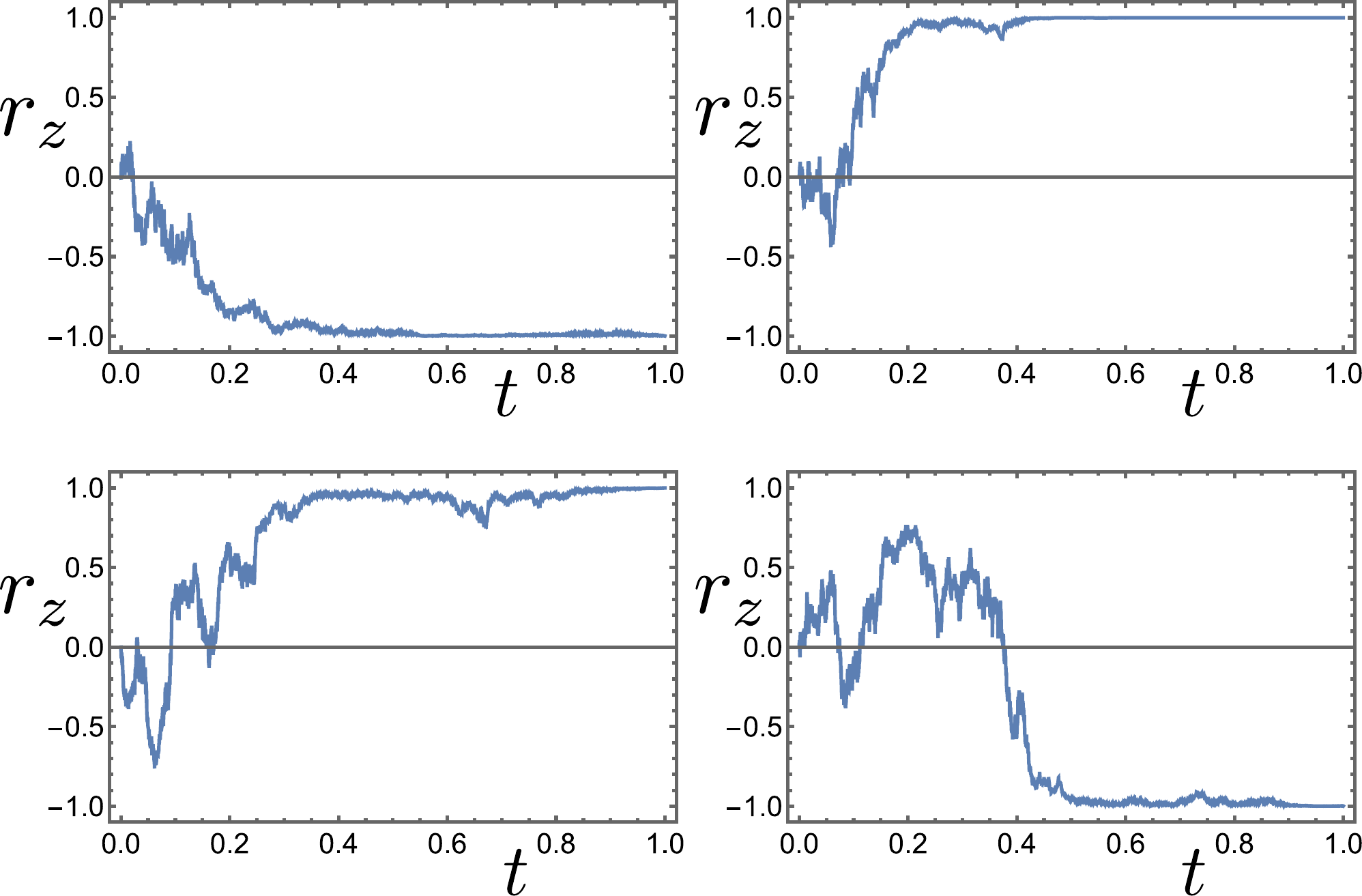}
\par\end{centering}
\caption{Four stochastic trajectories $r_{z}(t)$ derived from Eq. (\ref{eq:drz})
with strength of measurement $\alpha_{z}=1$. Starting at $r_{z}(0)=0$,
they evolve towards eigenstates of the $\sigma_{z}$ observable at
$r_{z}=\pm1$.\label{fig:Ensemble-of-trajectories.}}

\end{figure}

If we represent the density matrix in the form $\rho=\frac{1}{2}\left(\mathbb{I}+r_{z}\sigma_{z}\right)$,
it may be shown that the dynamics of Eq. (\ref{Ito}) correspond to
the evolution of the real stochastic variable $r_{z}(t)$ according
to \citep{jacobs2014quantum}
\begin{equation}
dr_{z}=2\alpha_{z}\left(1-r_{z}^{2}\right)dW.\label{eq:drz}
\end{equation}
Example realisations of such behaviour, starting from the fully mixed
state at $r_{z}(0)=0$, are shown in Fig. \ref{fig:Ensemble-of-trajectories.}.
Notice that $r_{z}$ evolves asymptotically towards $\pm1$, corresponding
to density matrices $\rho_{\pm}^{e}$, and note also that the average
increment $\langle dr_{z}\rangle$ over the ensemble satisfies $\langle dr_{z}\rangle=d\langle r_{z}\rangle=2\alpha_{z}\left(1-\langle r_{z}^{2}\rangle\right)\langle dW\rangle=0$,
implying that $\langle r_{z}\rangle$ is time independent and that
$\langle\rho\rangle$ is as well. A similar conclusion can be reached
simply by evaluating the right hand side of Eq. (\ref{Lindblad}). 

The standard Lindblad equation cannot capture system `collapse' to
an eigenstate, but instead describes the average behaviour of an ensemble
of collapsing systems. For a closer consideration of the dynamics
and thermodynamics of collapse, we need to `unravel' the standard
Lindblad equation into its stochastic version (\ref{Ito}), using
it to generate an ensemble of trajectories that model possible physical
evolutions of the open quantum system.

Using It\^o's lemma, it can be shown that the purity of the state,
$P={\rm Tr}\rho^{2}=\frac{1}{2}\left(1+r_{z}^{2}\right)$, evolves
according to
\begin{equation}
dP=8\alpha_{z}^{2}\left(1-P\right)^{2}dt+4\alpha_{z}r_{z}\left(1-P\right)dW.\label{eq:purity}
\end{equation}
The dynamics take the purity asymptotically towards a fixed point
at $P=1$, or the density matrix towards one of $\rho_{\pm}^{e}$,
which is clearly a natural consequence of the process of measurement.

\begin{figure}
\begin{centering}
\includegraphics[width=1\columnwidth]{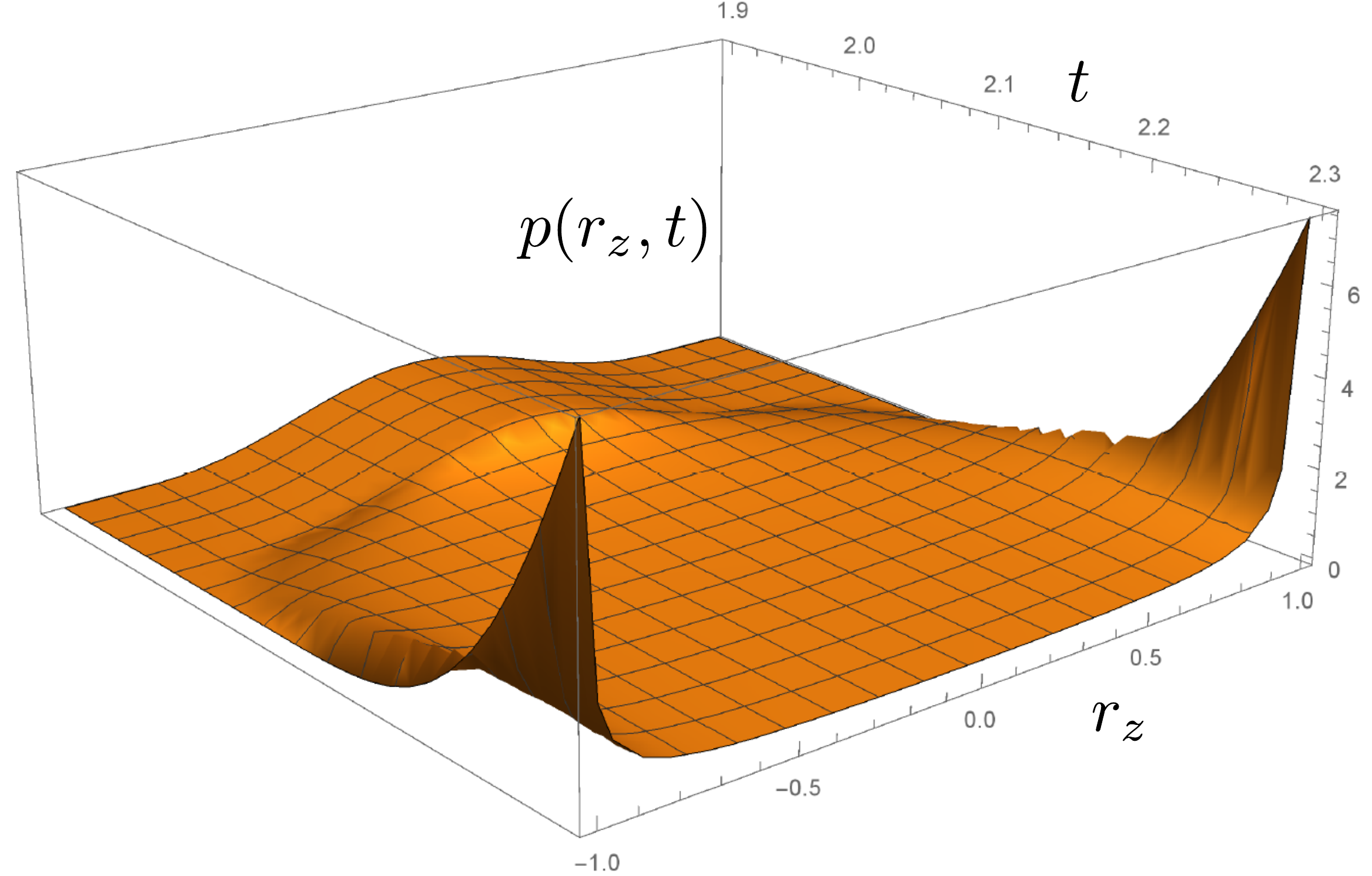}
\par\end{centering}
\caption{A probability density function $p(r_{z},t)$, evolving according to
the Fokker-Planck equation (\ref{FPE}), describing the evolution
of an ensemble of density matrices under measurement of $\sigma_{z}$.
A gaussian centred initially at the origin separates and probability
density accumulates asymptotically at $r_{z}=\pm1$. This approach
complements the direct computation of trajectories $r_{z}(t)$ illustrated
in Fig. \ref{fig:Ensemble-of-trajectories.}. \label{fig:FPE evolution}}
\end{figure}

The Fokker-Planck equation describing the evolution of the probability
density function (pdf) $p(r_{z},t)$ for the system variable $r_{z}$
is
\begin{equation}
\frac{\partial p}{\partial t}=\frac{\partial^{2}}{\partial r_{z}^{2}}\left(2\alpha_{z}^{2}\left(1-r_{z}^{2}\right)^{2}p\right),\label{FPE}
\end{equation}
and this provides further insight into the dynamics. Figure \ref{fig:FPE evolution}
illustrates the development starting from a gaussian pdf centred on
the maximally mixed state at $r_{z}=0$. The ensemble of density matrices
is separated by the dynamics into equal size groups that evolve asymptotically
towards the eigenstates of $\sigma_{z}$ at $r_{z}=\pm1$. The preservation
of the ensemble average of $r_{z}$ is apparent. 

\subsection{Stochastic entropy production}

The (total) stochastic entropy production associated with the evolution
of a stochastic variable in a certain time interval is defined in
terms of probabilities for the generation of a `forward' set of moves
in the phase space and the corresponding `backward' set \citep{seifert2008stochastic}.
For the coordinate $r_{z}$, and the time interval $dt$, we need
to consider the quantity
\begin{align}
 & d\Delta s_{{\rm tot}}(r_{z},t\to r_{z}+dr_{z},t+dt)\label{ds_tot}\\
 & =\ln\left({\rm Prob(forward)/Prob(backward)}\right)\nonumber \\
 & =\ln\frac{p(r_{z},t)\Delta r_{z}(r_{z})T(r_{z}\to r_{z}+dr_{z})}{p(r_{z}+dr_{z},t+dt)\Delta r_{z}(r_{z}+dr_{z})T(r_{z}+dr_{z}\to r_{z})},\nonumber 
\end{align}
where the $T$ are conditional probabilities for the transitions indicated.
For stochastic variables that are odd under time reversal symmetry,
additional features have to be included in this definition, but since
$r_{z}$ is even we can ignore such complications \citep{spinney2012entropy,Ford}.

It may be shown that the expectation or mean of $d\Delta s_{{\rm tot}}$
is never negative, which ultimately provides an underpinning for the
second law of thermodynamics \citep{seifert2008stochastic}.

We shall discuss the contributions to $d\Delta s_{{\rm tot}}$ involving
the pdf $p(r_{z},t)$ and the volume increment $\Delta r_{z}(r_{z})$
shortly, but first let us consider the ratio of conditional probabilities.
The two choices of forward move $\rho\to\rho^{\prime\pm}$ in Eqs.
(\ref{eq:moves}) and (\ref{Kraus}) are selected with probabilities
\begin{equation}
p_{\pm}=\frac{1}{2}\left(1\pm2\alpha_{z}r_{z}\sqrt{dt}\right).\label{p_=00005Cpm}
\end{equation}
The corresponding backward moves $\rho^{\prime\pm}\to\rho$ are described
by the quantum maps
\begin{equation}
\rho=\frac{\tilde{M}_{\mp}\rho^{\prime\pm}\tilde{M}_{\mp}^{\dagger}}{{\rm Tr}\left(\tilde{M}_{\mp}\rho^{\prime\pm}\tilde{M}_{\mp}^{\dagger}\right)},\label{Kraus backward}
\end{equation}
in terms of reverse Kraus operators $\tilde{M}_{\mp}$ that can be
identified from the condition that the initial density matrix is recovered.
Inserting Eq. (\ref{eq:moves}) into Eq. (\ref{Kraus backward}) we
have
\begin{equation}
\rho=\frac{\tilde{M}_{\mp}M_{\pm}\rho M_{\pm}^{\dagger}\tilde{M}_{\mp}^{\dagger}}{{\rm Tr}\left(\tilde{M}_{\mp}M_{\pm}\rho M_{\pm}^{\dagger}\tilde{M}_{\mp}^{\dagger}\right)},\label{Kraus condition}
\end{equation}
which requires $\tilde{M}_{\mp}M_{\pm}$ to be proportional to the
identity, up to linear order in $dt$. For $c=c^{\dagger}$ this can
be achieved using
\begin{equation}
\tilde{M}_{\mp}=\frac{1}{\sqrt{2}}\left(\mathbb{I}-\frac{1}{2}c^{2}dt\mp c\sqrt{dt}\right)=M_{\mp},\label{reverse Kraus}
\end{equation}
and specifically for $c=\alpha_{z}\sigma_{z}$ we have
\begin{equation}
\tilde{M}_{\mp}M_{\pm}=\frac{1}{2}\left(1-2\alpha_{z}^{2}dt\right)\mathbb{I}.\label{forward-backward Kraus product}
\end{equation}
Hence the probabilities for the backward moves are
\begin{equation}
p_{\mp}^{\prime}={\rm Tr}\left(\tilde{M}_{\mp}\rho^{\prime\pm}\tilde{M}_{\mp}^{\dagger}\right)=\frac{{\rm Tr}\left(M_{\mp}M_{\pm}\rho M_{\pm}^{\dagger}M_{\mp}^{\dagger}\right)}{{\rm Tr}\left(M_{\pm}\rho M_{\pm}^{\dagger}\right)},\label{p'_=00005Cmp}
\end{equation}
leading to 
\begin{equation}
p_{\mp}^{\prime}=\frac{\left(1-4\alpha_{z}^{2}dt\right)}{2\left(1\pm2\alpha_{z}r_{z}\sqrt{dt}\right)}.\label{eq:p'=00005Cmp}
\end{equation}
The ratio of conditional probabilities $T(r_{z}\to r_{z}+dr_{z}^{\pm})/T(r_{z}+dr_{z}^{\pm}\to r_{z})$
is then
\begin{equation}
\frac{p_{\pm}}{p_{\mp}^{\prime}}=1\pm4\alpha_{z}r_{z}\sqrt{dt}+4\alpha_{z}^{2}\left(1+r_{z}^{2}\right)dt.\label{p ratio}
\end{equation}

\begin{figure}
\begin{centering}
\includegraphics[width=1\columnwidth]{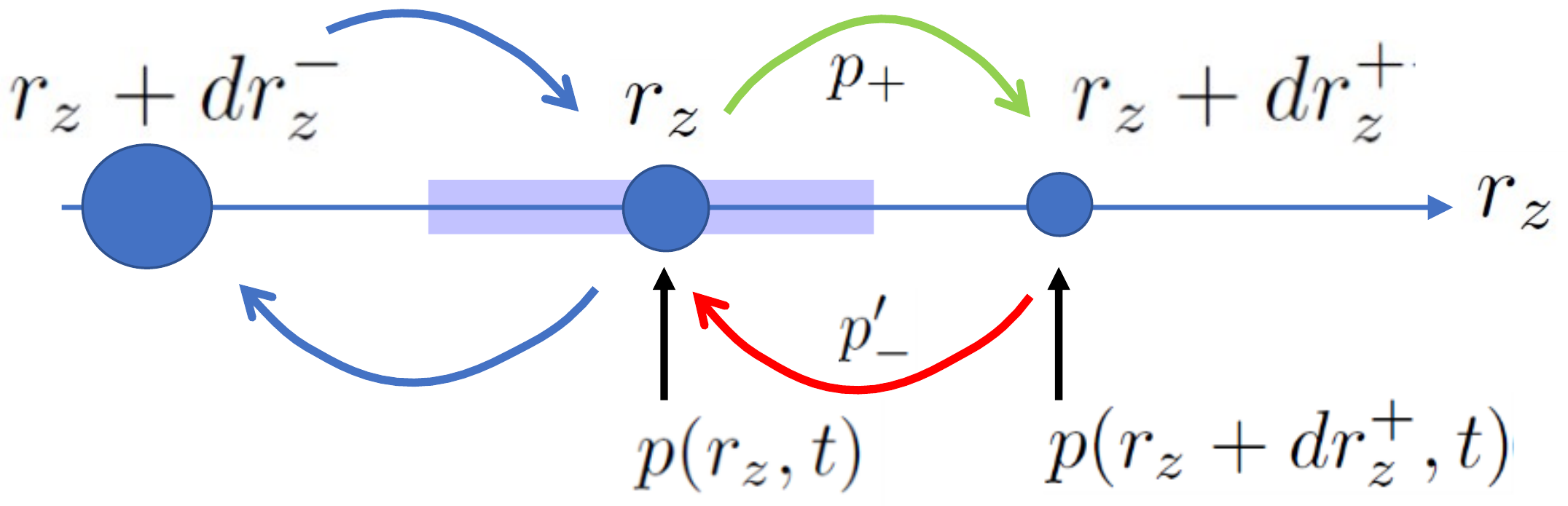}
\par\end{centering}
\caption{Available incremental moves on a set of locations on the $r_{z}$
axis according to the stochastic dynamics of measurement of $\sigma_{z}$,
illustrating Eqs. (\ref{p_=00005Cpm}), (\ref{eq:p'=00005Cmp}) and
(\ref{drz}). The size of the circles represents the local probability
density $p(r_{z},t)$. The shaded rectangle represents the volume
$\Delta r_{z}=\frac{1}{2}\left(dr_{z}^{+}-dr_{z}^{-}\right)$ of the
continuum phase space associated with a given location $r_{z}$. \label{fig:Moves-on-the}}
\end{figure}

The two possible increments in $r_{z}$ are
\begin{align}
dr_{z}^{\pm} & ={\rm Tr}\left(\rho^{\prime\pm}\sigma_{z}\right)-r_{z}\nonumber \\
 & =-4\alpha_{z}^{2}r_{z}\left(1-r_{z}^{2}\right)dt\pm2\alpha_{z}\left(1-r_{z}^{2}\right)\sqrt{dt},\label{drz}
\end{align}
and we note that the mean and variance over the two possibilities
are
\begin{equation}
\langle dr_{z}\rangle=p_{+}dr_{z}^{+}+p_{-}dr_{z}^{-}=0,\label{eq:average drz}
\end{equation}
and 
\begin{align}
\sigma_{r_{z}}^{2} & =p_{+}\left(dr_{z}^{+}-\langle dr_{z}\rangle\right)^{2}+p_{-}\left(dr_{z}^{-}-\langle dr_{z}\rangle\right)^{2}\nonumber \\
 & =4\alpha_{z}^{2}\left(1-r_{z}^{2}\right)^{2}dt.\label{eq:variance drz}
\end{align}
confirming that the evolution is consistent with the SDE for $r_{z}$
in Eq. (\ref{eq:drz}). The moves and their probabilities are illustrated
in Fig. \ref{fig:Moves-on-the}. 

We now write 
\begin{equation}
d\Delta s_{{\rm tot}}^{\pm}=d\Delta s_{A}^{\pm}+d\Delta s_{B}^{\pm},\label{dstot terms}
\end{equation}
where 
\begin{equation}
d\Delta s_{A}^{\pm}=\ln\left(\frac{T(r_{z}\to r_{z}+dr_{z}^{\pm})}{T(r_{z}+dr_{z}^{\pm}\to r_{z})}\right)=\ln\left(\frac{p_{\pm}}{p_{\mp}^{\prime}}\right),\label{eq:dsa}
\end{equation}
and
\begin{equation}
d\Delta s_{B}^{\pm}=\ln\left(\frac{p(r_{z},t)\Delta r_{z}(r_{z})}{p(r_{z}+dr_{z}^{\pm},t+dt)\Delta r_{z}(r_{z}+dr_{z}^{\pm})}\right).\label{dss}
\end{equation}
Inserting Eq. (\ref{p ratio}) we have
\begin{align}
 & d\Delta s_{A}^{\pm}=\pm4\alpha_{z}r_{z}\sqrt{dt}+4\alpha_{z}^{2}\left(1-r_{z}^{2}\right)dt,\label{ln pratio}
\end{align}
which provides two choices of incremental contribution to the stochastic
entropy production in the forward move. We can compute the mean of
$d\Delta s_{A}^{\pm}$: 
\begin{align}
\langle d\Delta s_{A}\rangle & =p_{+}d\Delta s_{A}^{+}+p_{-}d\Delta s_{A}^{-}\nonumber \\
 & =\left(p_{+}-p_{-}\right)4\alpha_{z}r_{z}\sqrt{dt}+\left(p_{+}+p_{-}\right)4\alpha_{z}^{2}\left(1-r_{z}^{2}\right)dt\nonumber \\
 & =4\alpha_{z}^{2}\left(1+r_{z}^{2}\right)dt,\label{mean dsc}
\end{align}
and the variance:
\begin{align}
\sigma_{A}^{2} & =p_{+}\left(d\Delta s_{A}^{+}-\langle d\Delta s_{A}\rangle\right)^{2}+p_{-}\left(d\Delta s_{A}^{-}-\langle d\Delta s_{A}\rangle\right)^{2}\nonumber \\
 & =16\alpha_{z}^{2}r_{z}^{2}dt,\label{variance c}
\end{align}
from which we conclude that the evolution can be represented by an
It\^o process for a stochastic variable $\Delta s_{A}$:
\begin{align}
d\Delta s_{A} & =4\alpha_{z}^{2}\left(1+r_{z}^{2}\right)dt+4\alpha_{z}r_{z}dW.\label{dsc SDE}
\end{align}

We next consider the contribution $d\Delta s_{B}^{\pm}$ to the stochastic
entropy production given in Eq. (\ref{dss}). The volume $\Delta r_{z}(r_{z})$
is the region bounded by increments $\frac{1}{2}dr_{z}^{\pm}$ starting
from $r_{z}$. It is the patch of phase space associated with coordinate
$r_{z}$, as illustrated in Fig. \ref{fig:Moves-on-the}. We write
$\Delta r_{z}=\frac{1}{2}\left(dr_{z}^{+}-dr_{z}^{-}\right)=2\alpha_{z}\left(1-r_{z}^{2}\right)\sqrt{dt}$
and then

\begin{equation}
d\Delta s_{B}^{\pm}=-d\ln p^{\pm}+d\Delta s_{C}^{\pm},\label{dss terms}
\end{equation}
where $d\ln p^{\pm}=\ln p(r_{z}+dr_{z}^{\pm},t+dt)-\ln p(r_{z},t)$
and 
\begin{align}
d\Delta s_{C}^{\pm} & =\ln\left(\frac{\Delta r_{z}(r_{z})}{\Delta r_{z}(r_{z}+dr_{z}^{\pm})}\right)\nonumber \\
 & =4\alpha_{z}^{2}\left(1-r_{z}^{2}\right)dt\pm4\alpha_{z}r_{z}\sqrt{dt}.\label{dsv}
\end{align}
The mean of $d\Delta s_{C}^{\pm}$ is 
\begin{align}
\langle d\Delta s_{C}\rangle & =p_{+}d\Delta s_{C}^{+}+p_{-}d\Delta s_{C}^{-}\nonumber \\
 & =4\alpha_{z}^{2}\left(1+r_{z}^{2}\right)dt,\label{mean dsv}
\end{align}
and the variance is
\begin{align}
\sigma_{C}^{2} & =p_{+}\left(d\Delta s_{C}^{+}-\langle d\Delta s_{C}\rangle\right)^{2}+p_{-}\left(d\Delta s_{C}^{-}-\langle d\Delta s_{C}\rangle\right)^{2}\nonumber \\
 & =16\alpha_{z}^{2}r_{z}^{2}dt,\label{variance v}
\end{align}
so the It\^o process for this component of stochastic entropy production
is
\begin{align}
d\Delta s_{C} & =4\alpha_{z}^{2}\left(1+r_{z}^{2}\right)dt+4\alpha_{z}r_{z}dW.\label{dsv SDE}
\end{align}

Similarly, it may be shown that the term $-d\ln p^{\pm}$ in Eq. (\ref{dss terms})
makes a contribution of $-d\ln p$ to the It\^o process for $d\Delta s_{{\rm tot}}$.
Combining this with Eqs. (\ref{dstot terms}), (\ref{dsc SDE}), (\ref{dss terms})
and (\ref{dsv SDE}), the  stochastic entropy production can be shown
to evolve according to the It\^o process
\begin{equation}
d\Delta s_{{\rm tot}}=-d\ln p(r_{z},t)+8\alpha_{z}^{2}\left(1+r_{z}^{2}\right)dt+8\alpha_{z}r_{z}dW.\label{dstot}
\end{equation}

The term $-d\ln p(r_{z},t)$ is usually referred to as the stochastic
entropy production of the system, $d\Delta s_{{\rm sys}}$. The remaining
terms are then regarded as stochastic entropy production in the environment
(in this case the measuring device), and denoted $d\Delta s_{{\rm env}}$
or $d\Delta s_{{\rm meas}}.$ Note that the evolution of the stochastic
entropy production in Eq. (\ref{dstot}), with a system contribution
that depends on the pdf $p(r_{z},t)$ over the phase space of the
density matrix, is continuous. This is in contrast to implementations
of stochastic entropy production in quantum mechanics that involve
the probability distribution over eigenstates of the measured operator
in the formalism, or that invoke projective measurements causing discontinuities
that are potentially infinite in magnitude \citep{elouard2017role}.

\subsection{Derivation of $d\Delta s_{{\rm tot}}$ from the dynamics}

The derivation of $d\Delta s_{{\rm tot}}$ in the previous section
is intricate, but there is an alternative approach that is much more
straightforward \citep{SpinneyFord12a,spinney2012entropy} and does
not require the identification of reverse Kraus operators \citep{crooks2008quantum}.
Let us consider an It\^o process for a stochastic variable $x$ in
the form
\begin{equation}
dx=\left(A^{{\rm rev}}(x,t)+A^{{\rm irr}}(x,t)\right)dt+B(x,t)dW,\label{dynamics SDE}
\end{equation}
where the terms proportional to $A^{{\rm rev}}$ and $A^{{\rm irr}}$
represent modes of deterministic dynamics that satisfy and violate
time reversal symmetry, respectively. Then the stochastic entropy
production is given by
\begin{eqnarray}
d\Delta s_{{\rm tot}} & = & -d\ln p(x,t)+\frac{A^{{\rm irr}}}{D}dx-\frac{A^{{\rm rev}}A^{{\rm irr}}}{D}dt+\frac{\partial A^{{\rm irr}}}{\partial x}dt\nonumber \\
 &  & -\frac{\partial A^{{\rm rev}}}{\partial x}dt-\frac{1}{D}\frac{\partial D}{\partial x}dx+\frac{(A^{{\rm rev}}-A^{{\rm irr}})}{D}\frac{\partial D}{\partial x}dt\nonumber \\
 &  & -\frac{\partial^{2}D}{\partial x^{2}}dt+\frac{1}{D}\left(\frac{\partial D}{\partial x}\right)^{2}dt,\label{dstot expression}
\end{eqnarray}
where $D(x,t)=\frac{1}{2}B(x,t)^{2}$. This expression might not seem
particularly intuitive, but for dynamics that possess a stationary
state with zero probability current, characterised by a pdf $p_{{\rm st}}(x)$,
Eq. (\ref{dstot expression}) reduces to the simpler expression $d\Delta s_{{\rm tot}}=-d\ln\left(p(x,t)/p_{{\rm st}}(x)\right)$,
and hence the stochastic entropy production is seen to arise from
deviation from stationarity.

For the dynamics of $r_{z}$ given by Eq. (\ref{eq:drz}) we have
$A^{{\rm rev}}=A^{{\rm irr}}=0$ and $B=2\alpha_{z}\left(1-r_{z}^{2}\right)$.
Hence $D=2\alpha_{z}^{2}(1-r_{z}^{2})^{2}$, leading to $dD/dr_{z}=-8\alpha_{z}^{2}r_{z}(1-r_{z}^{2})$,
$d^{2}D/dr_{z}^{2}=-8\alpha_{z}^{2}(1-3r_{z}^{2})$, and
\begin{align}
d\Delta s_{{\rm tot}} & =-d\ln p-\frac{1}{D}\frac{dD}{dr_{z}}dr_{z}-\frac{d^{2}D}{dr_{z}^{2}}dt+\frac{1}{D}\left(\frac{dD}{dr_{z}}\right)^{2}dt\nonumber \\
 & =-d\ln p+8\alpha_{z}^{2}\left(1+r_{z}^{2}\right)dt+8\alpha_{z}r_{z}dW.\label{dstot direct}
\end{align}
This is in agreement with Eq. (\ref{dstot}), but the derivation is
much more direct. Extension to sets of coupled It\^o processes for
several stochastic variables $\{x_{i}\}$ is straightforward, and
we shall encounter an example of such a generalisation in Section
\ref{sec:Simultaneous-measurement-of}.

\subsection{Results}

\begin{figure}
\begin{centering}
\includegraphics[width=1\columnwidth]{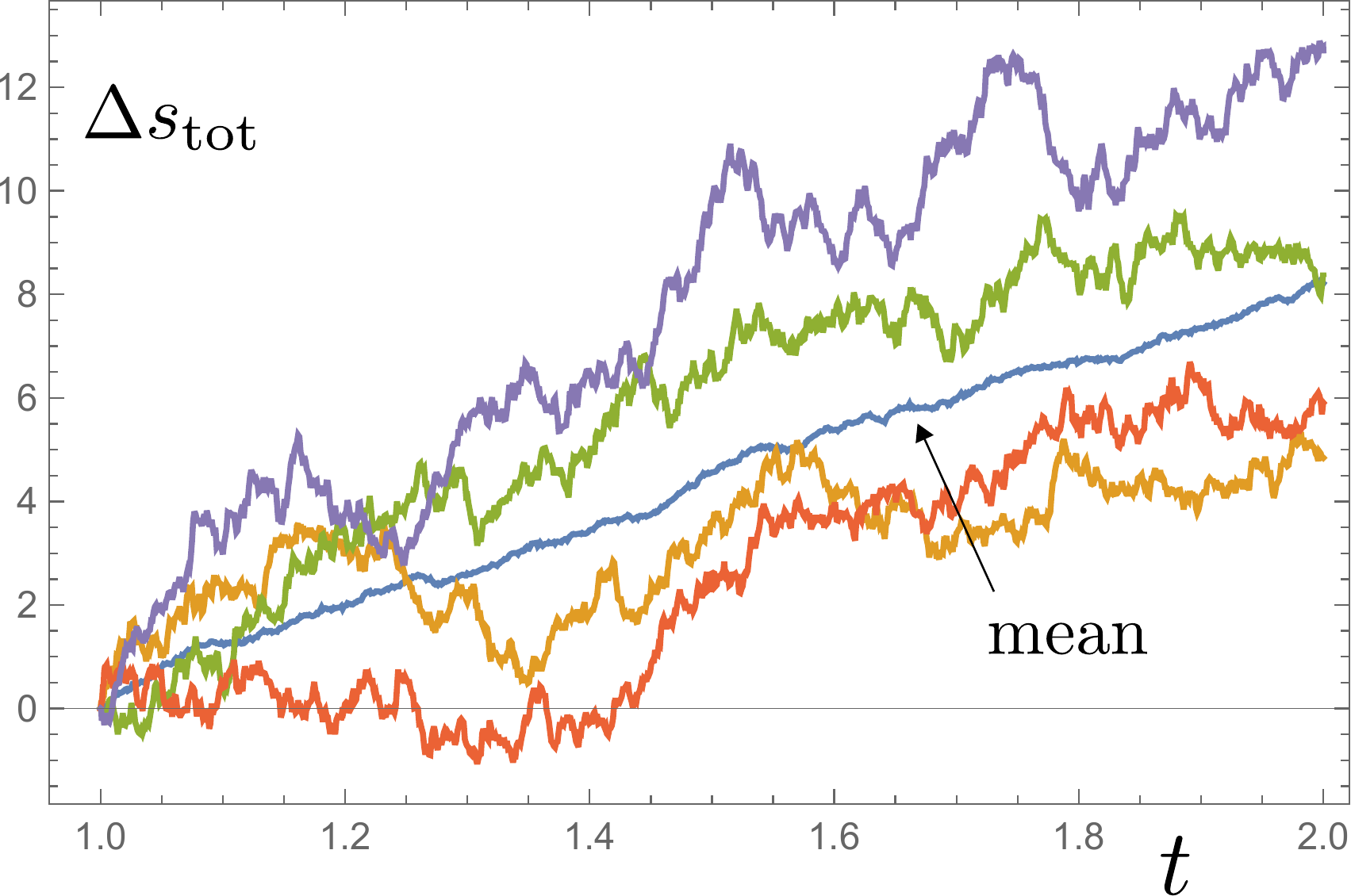}
\par\end{centering}
\caption{Four trajectories illustrating the stochastic entropy production $\Delta s_{{\rm tot}}(t)$
for the dynamics of Eq. (\ref{eq:drz}) in the interval $1\le t\le2$,
starting from a gaussian pdf centred on $r_{z}=0$ at $t=0$, and
with $\alpha_{z}=1$. The mean over a sample of 40 trajectories is
consistent with an asymptotic average rate of production equal to
$8\alpha_{z}^{2}$, as suggested in Eq. (\ref{mean asymptotic dstot}).\label{fig:s trajectories}}
\end{figure}

Let us now consider the character of the stochastic entropy production
described by Eq. (\ref{dstot direct}). It is straightforward to evaluate
$\Delta s_{{\rm tot}}(t)$ numerically, employing solutions to the
Fokker-Planck equation (\ref{FPE}) and the It\^o process for $r_{z}(t)$.
Example evolutions of $\Delta s_{{\rm tot}}(t)$ associated with trajectories
$r_{z}(t)$ are shown in Fig. \ref{fig:s trajectories}, for $\alpha_{z}=1$.
The mean stochastic entropy production over a sample of trajectories
appears to rise linearly in time. The increase reflects the fact that
the pdf $p(r_{z},t)$ does not reach a stationary state, but instead
progressively sharpens towards two $\delta$-function peaks at $r_{z}=\pm1$.
The system approaches one of the eigenstates but does not reach it
in finite time. A system that continues to evolve in response to time
reversal asymmetric dynamics (which includes the noise term as well
as the deterministic contribution proportional to $A^{{\rm irr}}$
in Eq. (\ref{dynamics SDE})) is characterised by stochastic entropy
production. 

The calculations of $\Delta s_{{\rm tot}}$ in Fig. \ref{fig:s trajectories}
were obtained after performing a transformation of the stochastic
variable to avoid difficulties arising from the singularities in $p(r_{z},t)$
as $t\to\infty$. It is possible to do this since the stochastic entropy
production is invariant under a coordinate transformation. Consider,
then, the variable $y=\tanh^{-1}r_{z}$, which evolves in time according
to
\begin{equation}
dy=4\alpha_{z}^{2}\tanh y\,dt+2\alpha_{z}dW,\label{dy}
\end{equation}
using It\^o's lemma. The phase space $-1\le r_{z}\le1$ maps to $-\infty\le y\le\infty$.
We identify $A^{{\rm rev}}(y)=0$, $A^{{\rm irr}}(y)=4\alpha_{z}^{2}\tanh y$,
$D(y)=2\alpha_{z}^{2}$ and write
\begin{align}
 & d\Delta s_{{\rm tot}}=-d\ln p(y,t)+\frac{A^{{\rm irr}}}{D}dy+\frac{dA^{{\rm irr}}}{dy}dt\nonumber \\
 & =-d\ln p(y,t)+4\alpha_{z}^{2}\left(1+\tanh^{2}y\right)dt+4\alpha_{z}\tanh y\,dW,\label{eq:dstot y}
\end{align}
where the pdf for $y$ satisfies the Fokker-Planck equation
\begin{equation}
\frac{\partial p}{\partial t}=-4\alpha_{z}^{2}\frac{\partial}{\partial y}\left(\tanh y\,p\right)+2\alpha_{z}^{2}\frac{\partial^{2}p}{\partial y^{2}}.\label{fpe y}
\end{equation}
Solving Eqs. (\ref{dy}), (\ref{eq:dstot y}) and (\ref{fpe y}) numerically
produces the trajectories in Fig. \ref{fig:s trajectories}. 

We can perform an analysis of the evolution at late times, where $r_{z}$
is close to 1 or $-1$ such that $\vert y\vert$ is large. The dynamics
are then approximated by
\begin{equation}
dy=\pm4\alpha_{z}^{2}dt+2\alpha_{z}dW,\label{asymptotic dy}
\end{equation}
employing the plus sign if $y>0$ and the negative if $y<0$. The
Fokker-Planck equation is
\begin{equation}
\frac{\partial p}{\partial t}=-4\alpha_{z}^{2}{\rm sgn}(y)\frac{\partial p}{\partial y}+2\alpha_{z}^{2}\frac{\partial^{2}p}{\partial y^{2}},\label{asymptotic fpe}
\end{equation}
which has an approximate asymptotic solution: 
\begin{equation}
p(y,t)\!\propto\!\frac{1}{t^{1/2}}\!\left[\exp\left[-\frac{(y-4\alpha_{z}^{2}t)^{2}}{8\alpha_{z}^{2}t}\right]+\exp\left[-\frac{(y+4\alpha_{z}^{2}t)^{2}}{8\alpha_{z}^{2}t}\right]\right],\label{asymptotic f(y)}
\end{equation}
consisting of two gaussians in the $y$ phase space, drifting with
equal and opposite velocities towards $\pm\infty$ and simultaneously
broadening. 

From Eq. (\ref{eq:dstot y}) we obtain stochastic entropy production
for a trajectory with $y\gg0$ of
\begin{equation}
d\Delta s_{{\rm tot}}\approx-d\ln p_{+}(y,t)+8\alpha_{z}^{2}dt+4\alpha_{z}\,dW,\label{asymptotic dstot}
\end{equation}
with 
\begin{equation}
p_{+}\propto\frac{1}{t^{1/2}}\exp\left(-\frac{(y-4\alpha_{z}^{2}t)^{2}}{8\alpha_{z}^{2}t}\right),\label{f+}
\end{equation}
and hence
\begin{equation}
d\Delta s_{{\rm tot}}\approx d\left(\frac{(y-4\alpha_{z}^{2}t)^{2}}{8\alpha_{z}^{2}t}\right)+\frac{1}{2}d\ln t+8\alpha_{z}^{2}dt+4\alpha_{z}\,dW,\label{asymptotic dstot 2}
\end{equation}
the average of which is
\begin{align}
d\langle\Delta s_{{\rm tot}}\rangle & \approx\frac{1}{t}dt-\frac{\langle(y-4\alpha_{z}^{2}t)^{2}\rangle}{8\alpha_{z}^{2}t^{2}}dt+8\alpha_{z}^{2}dt\nonumber \\
 & =\frac{1}{t}dt-\frac{4\alpha^{2}t}{8\alpha^{2}t^{2}}dt+8\alpha_{z}^{2}dt,\label{mean asymptotic dstot}
\end{align}
which reduces to $8\alpha_{z}^{2}dt$ as $t\to\infty$. A similar
conclusion can be reached if $y\ll0$, so we expect mean stochastic
entropy production at a constant rate $8\alpha_{z}^{2}$ as $t\to\infty$,
confirming the behaviour seen in Fig. \ref{fig:s trajectories}.

\subsection{Contrast with von Neumann entropy \label{subsec:Contrast-with-von}}

At this point we should consider whether stochastic entropy production
is related to a change in the von Neumann entropy $S_{{\rm vN}}=-{\rm Tr}\rho\ln\rho$,
a commonly employed expression for entropy in quantum mechanics. 

The mean stochastic entropy production is the change in subjective
uncertainty with regard to the quantum state adopted by the world.
We are unable to make exact predictions when the dynamical influence
of the environment on the system is not specified in detail. The dynamics
then become effectively stochastic and our knowledge of the adopted
state is reduced with time.

In contrast, the von Neumann entropy is the uncertainty inherent to
a quantum state with regard to the outcomes of projective measurement
in a basis in which the density matrix is diagonal. It is a Shannon
entropy $-\sum_{i}P_{i}\ln P_{i}$ where $P_{i}$ is the probability
of projection into eigenstate $i$ of the observable. For a two level
system the number of such outcomes is two and so the von Neumann entropy
has an upper limit of $\ln2$. 

In contrast, the upper limit of the mean stochastic entropy production,
representing the change in uncertainty in the adopted quantum state
of the world, is infinite, since there is a continuum of possible
states that could be taken. The continued mean production of stochastic
entropy associated with measurement, discussed in previous sections,
represents this progressively greater uncertainty.

Note also that the stochastic entropy production we have been considering
has no connection with heat transfer or work. The two level system
under consideration does not possess a Hamiltonian $H$ and the adoption
of one or other level as a result of measurement does not involve
a change in system energy: specifically ${\rm Tr}H\rho=0$ throughout.
Stochastic entropy production is not necessarily associated with the
dissipation of potential energy into heat. Indeed it need not be in
classical mechanics, for example in the free expansion of an ideal
gas. In both classical and quantum settings the purpose of entropy
is to specify the degree of configurational uncertainty of a system.
In classical mechanics the configurations are described by sets of
classical coordinates: in quantum mechanics they are specified by
collections of (reduced) density matrix elements.

\begin{figure}
\begin{centering}
\includegraphics[width=1\columnwidth]{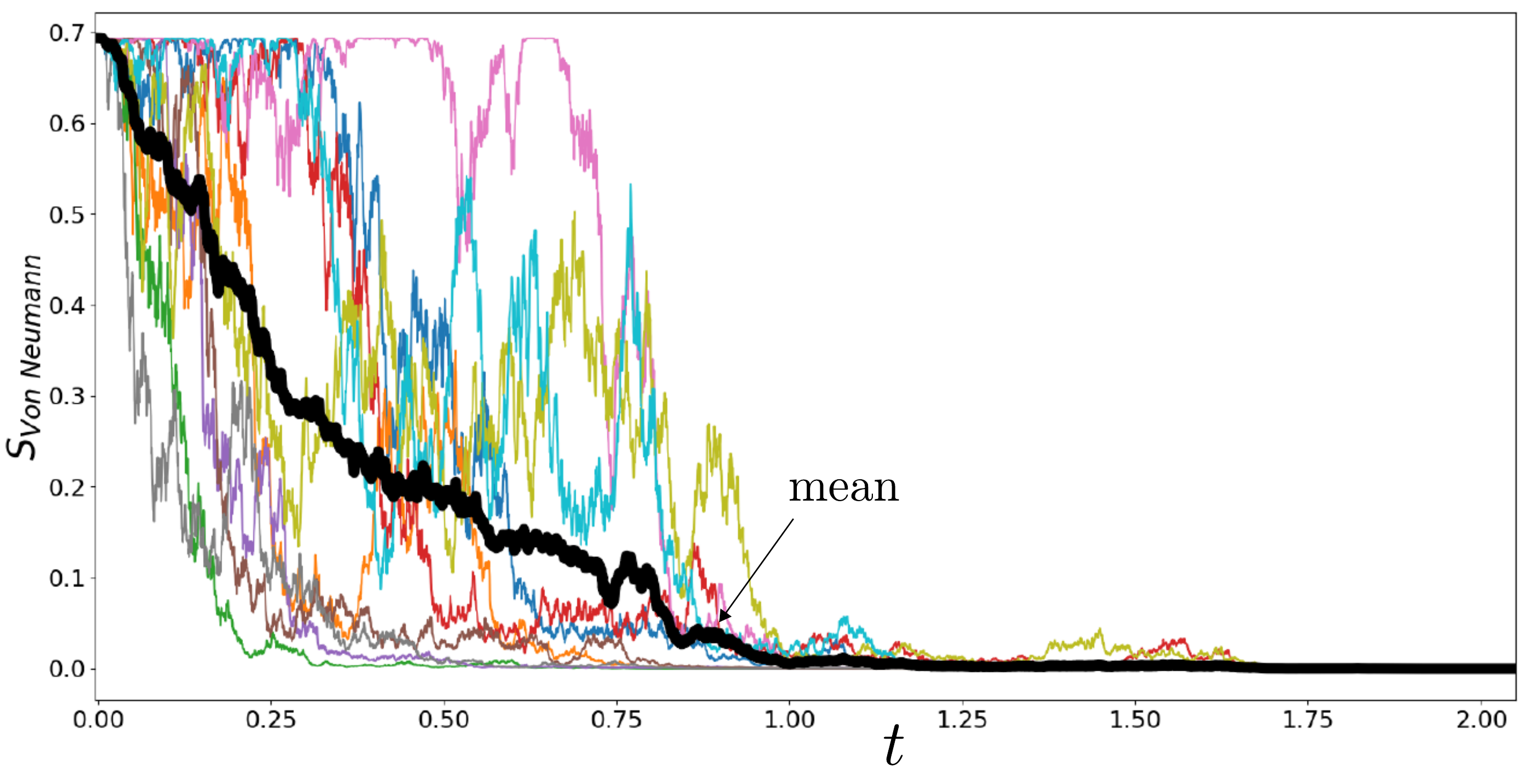}
\par\end{centering}
\caption{Evolution of the von Neumann entropy of the reduced density matrix
of the two level system, for 10 stochastic trajectories governed by
the dynamics of Eq. (\ref{eq:drz}) with $\alpha_{z}=1$. Mean behaviour
is also shown. Asymptotic values of zero imply that the system is
purified. \label{fig:von-Neumann-entropy}}

\end{figure}

Von Neumann entropy does play a role in computing the thermodynamic
entropy of a quantum system in a situation where it is subjected to
projective measurement and thereafter regarded as occupying one of
the eigenstates. However, it is not straightforward to employ von
Neumann entropy in discussions of the second law and the arrow of
time. The first issue is that the von Neumann entropy $-{\rm Tr}\bar{\rho}\ln\bar{\rho}$
of the ensemble averaged density matrix $\bar{\rho}$ remains constant
under the measurement dynamics employed here (because $\bar{\rho}$
remains constant). In contrast, the von Neumann entropy of a typical
member of the considered ensemble of density matrices falls to zero
under the dynamics. This is illustrated in Fig. \ref{fig:von-Neumann-entropy}
for the two level system where $\rho$ evolves towards one of the
$\rho_{\pm}^{e}$: the latter are pure states with $S_{{\rm vN}}=0$.
The mean von Neumann entropy change $-\Delta{\rm Tr}\langle\rho\ln\rho\rangle$
associated with the measurement process is then negative. In order
to protect the second law we need to consider entropy change in the
environment. The total stochastic entropy production production includes
such a contribution and so provides a more inclusive framework for
discussions of irreversibility.

\section{Simultaneous measurement of $\sigma_{z}$ and $\sigma_{x}$\label{sec:Simultaneous-measurement-of}}

\subsection{Evolution towards purity}

Now we turn our attention to a more complicated case of stochastic
entropy production associated with the dynamics of an open quantum
system. We continue to use the framework of quantum state diffusion,
involving transformations according to Eq. (\ref{eq:moves}), but
we now represent the stochastic influence of the environment on the
system using \emph{two} pairs of Kraus operators, given by
\begin{align}
M_{1\pm} & =\frac{1}{2}\left(\mathbb{I}-\frac{1}{2}c_{1}^{\dagger}c_{1}dt\pm c_{1}\sqrt{dt}\right)\nonumber \\
M_{2\pm} & =\frac{1}{2}\left(\mathbb{I}-\frac{1}{2}c_{2}^{\dagger}c_{2}dt\pm c_{2}\sqrt{dt}\right),\label{eq:two pairs of Kraus}
\end{align}
with $c_{1}=\alpha_{z}\sigma_{z}$ and $c_{2}=\alpha_{x}\sigma_{x}$.
The first and second pair describe the dynamics of continuous measurement
of observables $\sigma_{z}$ and $\sigma_{x}$, respectively, and
together therefore represent the performance of simultaneous measurement.
Since $\sigma_{z}$ and $\sigma_{x}$ do not commute, we expect this
not to result in a fixed outcome, and quantum state diffusion provides
an interesting illustration of what this means. 

\begin{figure}
\begin{centering}
\includegraphics[width=1\columnwidth]{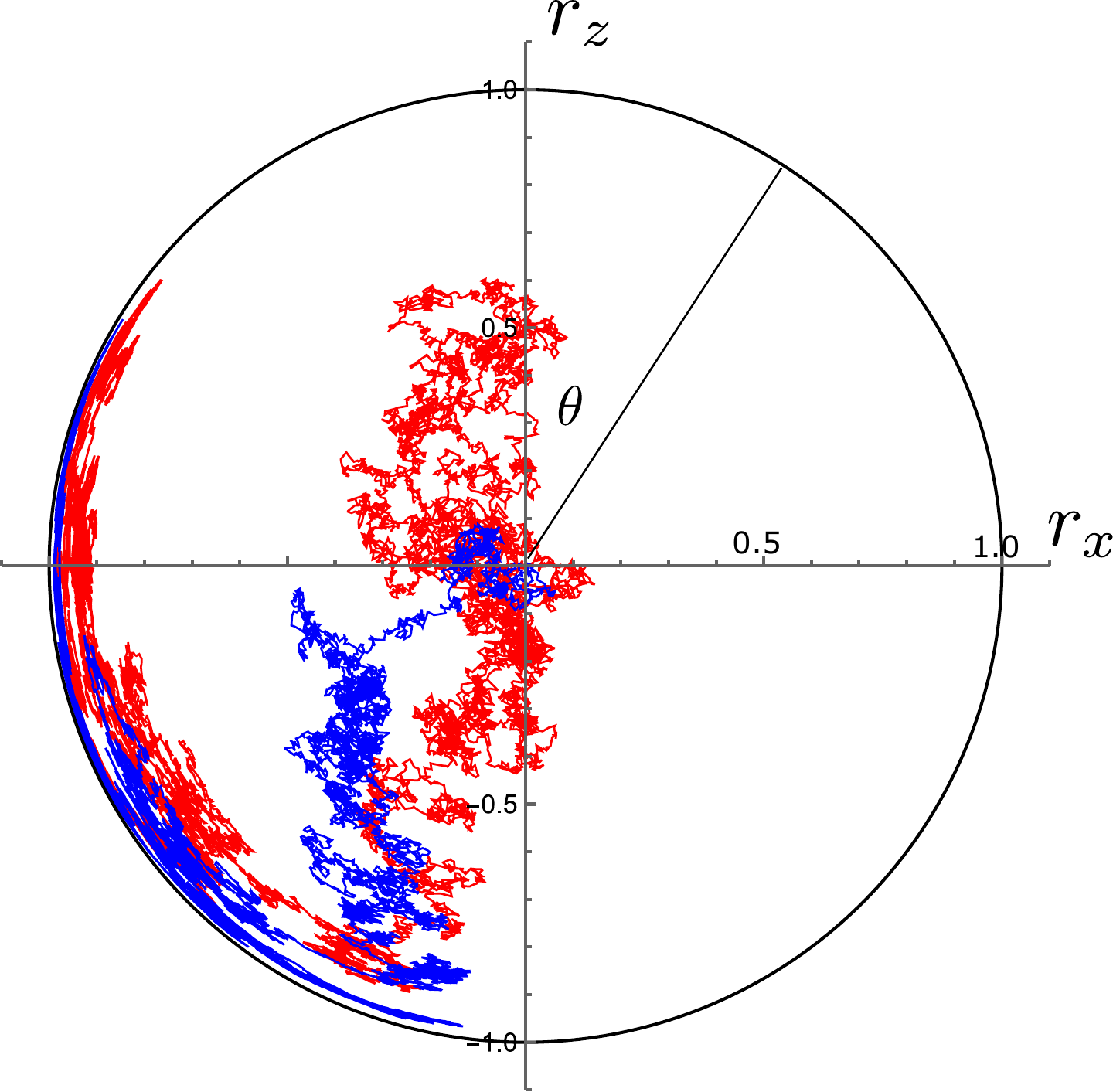}
\par\end{centering}
\caption{Two trajectories of the density matrix coordinates ($r_{x}(t),r_{z}(t)$)
generated by the dynamics of simultaneous measurement of $\sigma_{x}$
and $\sigma_{z}$, Eq. (\ref{two SDEs}), starting from the maximally
mixed state at the origin and for equal strengths of measurement $\alpha_{x}$
and $\alpha_{z}$. The outer black circle represents a condition of
purity, towards which the system evolves. Eigenstates of $\sigma_{x}$
and $\sigma_{z}$ lie at $\theta=\pm\pi/2$ and $\theta=0,\pi$ on
the circle, respectively.\label{fig:2d-trajectory.}}

\end{figure}

\begin{figure}
\centering{}\includegraphics[width=1\columnwidth]{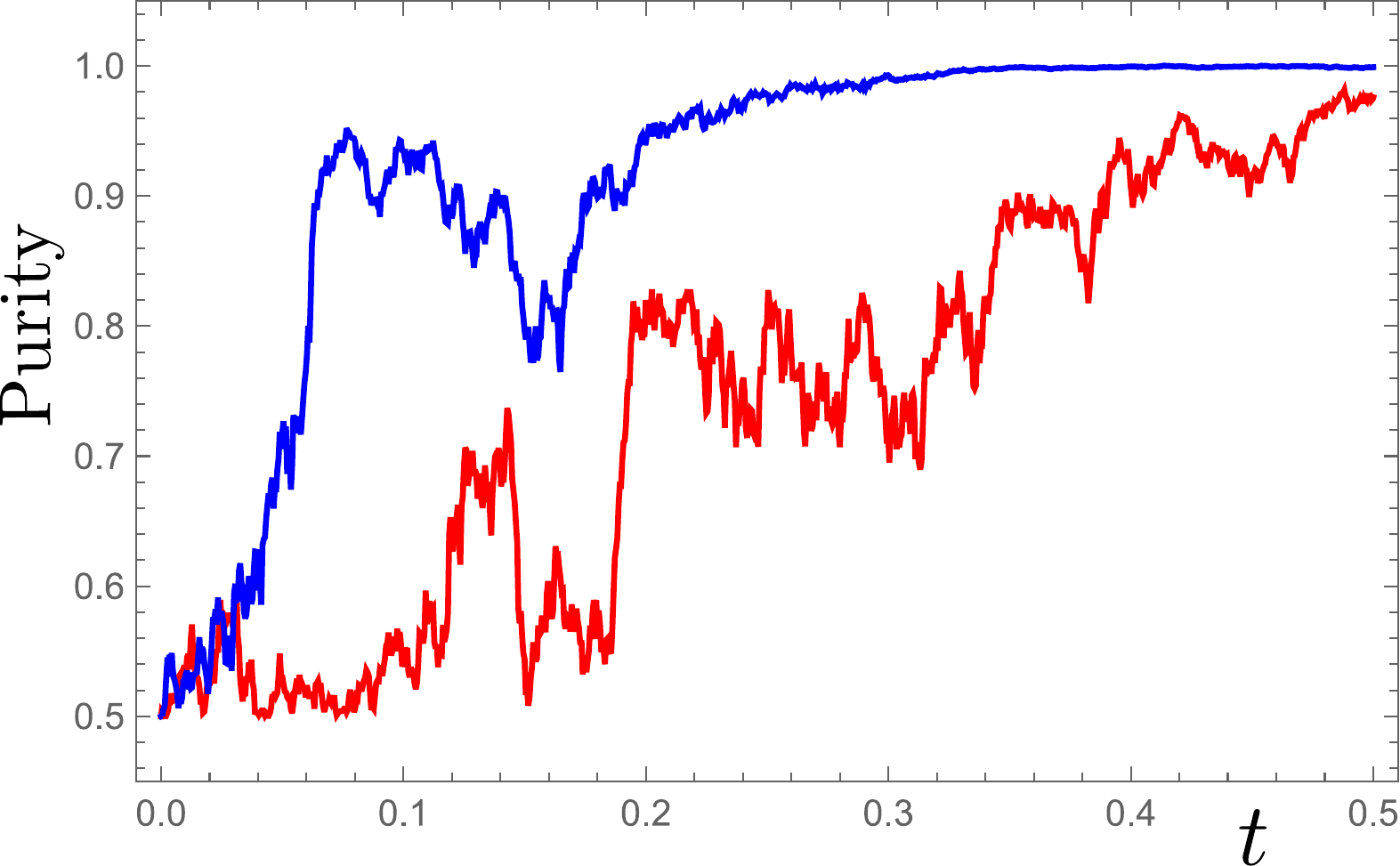}\caption{Evolution of purity for the system trajectories in Fig. \ref{fig:2d-trajectory.}.\label{fig:Evolution-of-purity}}
\end{figure}

Probabilities of stochastic changes in the reduced density matrix
of the system, brought about by interactions with the environment,
may be deduced for these operators, and a stochastic Lindblad equation
for its evolution may be derived:
\begin{align}
d\rho & =\sum_{i=1,2}\left(c_{i}\rho c_{i}^{\dagger}-\frac{1}{2}\rho c_{i}^{\dagger}c_{i}-\frac{1}{2}c_{i}^{\dagger}c_{i}\rho\right)dt\nonumber \\
 & \qquad+\left(\rho c_{i}^{\dagger}+c_{i}\rho-C_{i}\rho\right)dW_{i},\label{two pairs stochastic lindblad}
\end{align}
with $C_{i}={\rm Tr}\left((c_{i}+c_{i}^{\dagger})\rho\right)$. Upon
inserting the representation $\rho=\frac{1}{2}\left(\mathbb{I}+r_{z}\sigma_{z}+r_{x}\sigma_{x}\right)$,
the dynamics can be expressed as
\begin{align}
dr_{z} & =2\alpha_{z}\left(1-r_{z}^{2}\right)dW_{z}-2\alpha_{x}^{2}r_{z}dt-2\alpha_{x}r_{z}r_{x}dW_{x}\nonumber \\
dr_{x} & =2\alpha_{x}\left(1-r_{x}^{2}\right)dW_{x}-2\alpha_{z}^{2}r_{x}dt-2\alpha_{z}r_{x}r_{z}dW_{z},\label{two SDEs}
\end{align}
where $dW_{x}$ and $dW_{z}$ are independent Wiener increments.
Example stochastic trajectories starting from the maximally mixed
state at $r_{x}=r_{z}=0$ are shown in Fig. \ref{fig:2d-trajectory.}.
The purity $P={\rm Tr}\rho^{2}=\frac{1}{2}\left(1+r^{2}\right)$,
where $r^{2}=r_{x}^{2}+r_{z}^{2}$, evolves according to
\begin{align}
dP & =4\left(\alpha_{x}^{2}\left(1-r_{x}^{2}\right)+\alpha_{z}^{2}\left(1-r_{z}^{2}\right)\right)\left(1-P\right)dt\nonumber \\
 & +4\alpha_{x}r_{x}\left(1-P\right)dW_{x}+4\alpha_{z}r_{z}\left(1-P\right)dW_{z},\label{eq:2d purity}
\end{align}
such that $P=1$ is a fixed point reached asymptotically in time.
Examples of such system purification are shown in Fig. \ref{fig:Evolution-of-purity}.


The dynamics can be recast in terms of $Y=\tanh^{-1}r^{2}$, which
tends to $\infty$ as $r\to1$, and an angle $\theta=\tan^{-1}(r_{x}/r_{z})$.
For $\alpha_{x}=\alpha_{z}=\alpha$ the SDEs for these variables are
\begin{align}
dY & =\frac{4\alpha^{2}}{(1+\tanh Y)^{2}}\left(2+\tanh Y+3\tanh^{2}Y\right)dt\nonumber \\
 & \qquad+\frac{4\alpha\sqrt{\tanh Y}}{1+\tanh Y}dW_{Y}\nonumber \\
d\theta & =2\alpha\,dW_{\theta}/\sqrt{\tanh Y},\label{Y theta}
\end{align}
where $dW_{Y}=r^{-1}\left(r_{z}dW_{z}+r_{x}dW_{x}\right)$ and $dW_{\theta}=r^{-1}\left(-r_{x}dW_{z}+r_{z}dW_{x}\right)$
are independent Wiener increments. As $t\to\infty$, Eq. (\ref{eq:2d purity})
implies that $r^{2}\to1$ and hence $\tanh Y\to1$, in which case
we can write
\begin{equation}
dY\approx6\alpha^{2}dt+2\alpha\,dW_{Y},\label{asymptotic y theta}
\end{equation}
and so for late times we have $Y\approx6\alpha^{2}t+2\alpha W_{Y}+{\rm const}$.
The SDE for $\theta$ in this limit is $d\theta=2\alpha dW_{\theta}$,
such that the pdf becomes uniform over $\theta$. We write $p(Y,\theta,t)\to(2\pi)^{-1}F(Y,t)$,
in terms of a travelling and broadening gaussian in $Y$:
\begin{equation}
F(Y,t)=\frac{1}{\left(8\pi\alpha^{2}t\right)^{1/2}}\exp\left[-\frac{(Y-6\alpha^{2}t)^{2}}{8\alpha^{2}t}\right].\label{asymptotic y pdf}
\end{equation}

The stochastic entropy production can now be computed using the framework
of $Y$ and $\theta$ coordinates. We shall do so first for late times
where $Y\to1$ and the dynamical equations (\ref{Y theta}) become
independent. We can identify coefficients $A_{Y}^{{\rm irr}}=6\alpha^{2}$,
$A_{Y}^{{\rm rev}}=0$, $D_{Y}=2\alpha^{2}$, and $A_{\theta}^{{\rm irr}}=0$,
$A_{\theta}^{{\rm rev}}=0$, $D_{\theta}=2\alpha^{2}$ and use Eq.
(\ref{dstot expression}) to identify contributions to the stochastic
entropy production. The system stochastic entropy production can be
computed using the pdf in Eq. (\ref{asymptotic y pdf}). After some
manipulation we find that
\begin{equation}
d\Delta s_{{\rm tot}}\approx18\alpha^{2}dt+6\alpha\,dW_{Y},\label{eq:2d entropy}
\end{equation}
and thus the stochastic entropy production increases at a mean rate
of $18\alpha^{2}$. This is more than twice the mean rate of production
in Eq. (\ref{asymptotic dstot 2}) for the case of measurement of
$\sigma_{z}$ alone. The continued increase is once again a consequence
of the non-stationary character of the evolution: the dynamics have
the effect of purifying the system, but only as $t\to\infty.$

\begin{figure}
\begin{centering}
\includegraphics[width=1\columnwidth]{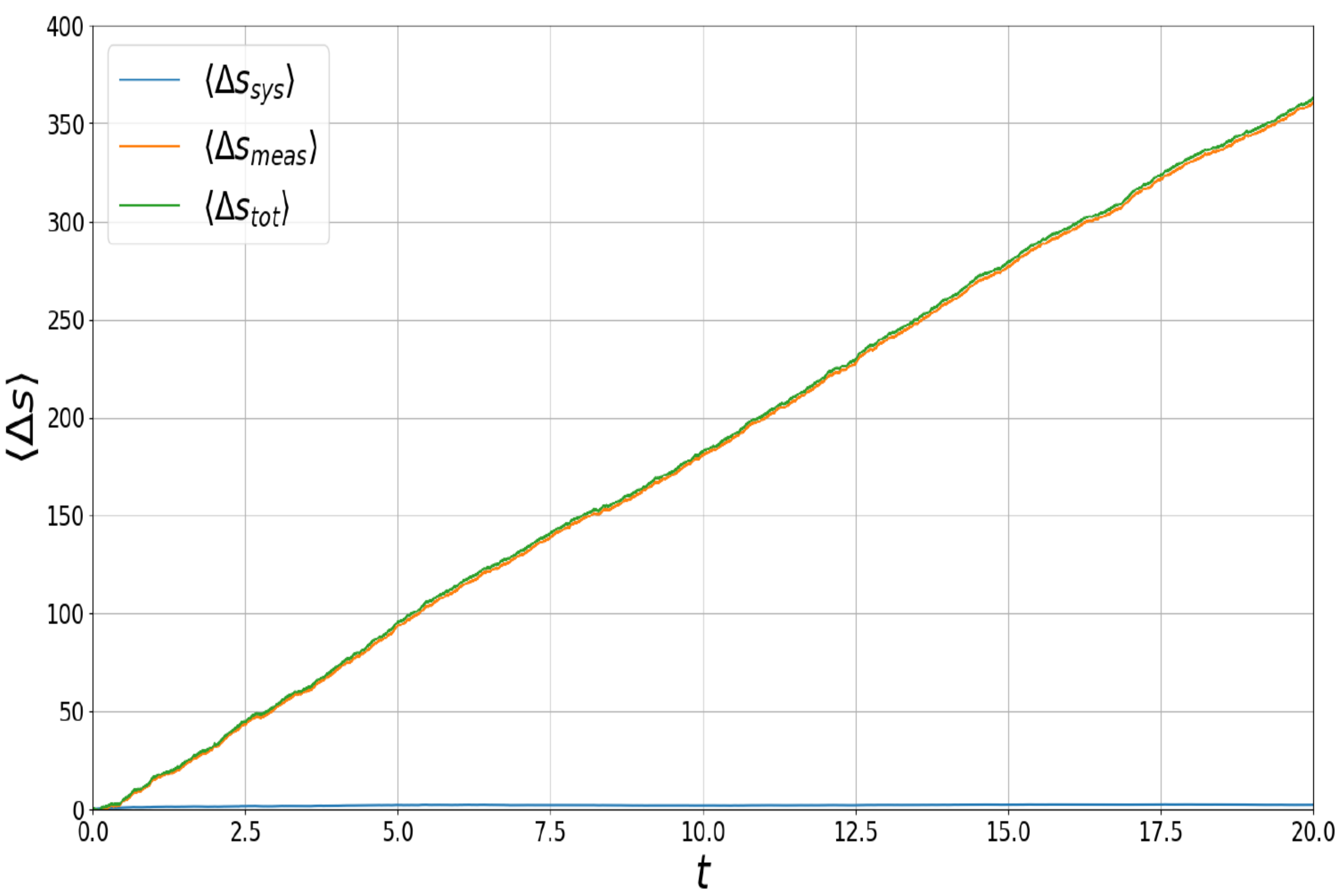}
\par\end{centering}
\caption{Mean stochastic entropy production $\langle\Delta s_{{\rm tot}}\rangle$
for simultaneous measurement of observables $\sigma_{x}$ and $\sigma_{z}$,
separated into contributions associated with the system and measuring
device, $\langle\Delta s_{{\rm sys}}\rangle$ and $\langle\Delta s_{{\rm meas}}\rangle$,
respectively. The strengths of measurement $\alpha_{x}$ and $\alpha_{z}$
are both set to unity and the numerically generated ensemble is comprised
of ten trajectories. The mean stochastic entropy production is consistent
with the estimate in Eq. (\ref{eq:2d entropy}). \label{fig:-for-2d}}

\end{figure}

For the more general situation, without taking $t$ to be large, it
is possible to compute the stochastic entropy production numerically,
based on the more elaborate coefficients of the SDEs in Eqs. (\ref{Y theta}),
and a general solution to the associated Fokker-Planck equation. Mean
stochastic entropy production over an ensemble of 10 trajectories
is given in Fig. \ref{fig:-for-2d}, separating $\langle\Delta s_{{\rm tot}}\rangle$
into contributions $\langle\Delta s_{{\rm sys}}\rangle=-\Delta\langle\ln p\rangle$
and $\langle\Delta s_{{\rm meas}}\rangle=\langle\Delta s_{{\rm tot}}\rangle-\langle\Delta s_{{\rm sys}}\rangle$.
The significance of this separation is that 
\begin{align}
-\Delta\langle\ln p\rangle & =-\int p(Y,\theta,t)\ln p(Y,\theta,t)dYd\theta\nonumber \\
 & +\int p(Y,\theta,0)\ln p(Y,\theta,0)dYd\theta,\label{eq:Delta S_G}
\end{align}
is the change in Gibbs entropy $\Delta S_{G}$ of the system when
described using the pdf in $Y,\theta$ coordinates. Note that the
Gibbs entropy is coordinate frame dependent and is therefore a measure
of the uncertainty of adopted state in a specific coordinate system.
In contrast, the mean stochastic entropy production is independent
of coordinate frame.

\subsection{Measurement of two non-commuting observables for a pure state}

Simultaneous measurement of $\sigma_{z}$ and $\sigma_{x}$ leads
asymptotically to a pure state located on a circle of radius $r=\sqrt{r_{x}^{2}+r_{z}^{2}}=1$
in the $(r_{x},r_{z})$ coordinate space. It is of interest now to
consider how the pdf over the angle $\theta$ (shown in Fig. \ref{fig:2d-trajectory.})
depends on the relative strengths of measurement of the two observables,
and to compute the stochastic entropy production arising from changes
in this ratio. 

\begin{figure}
\begin{centering}
\includegraphics[width=1\columnwidth]{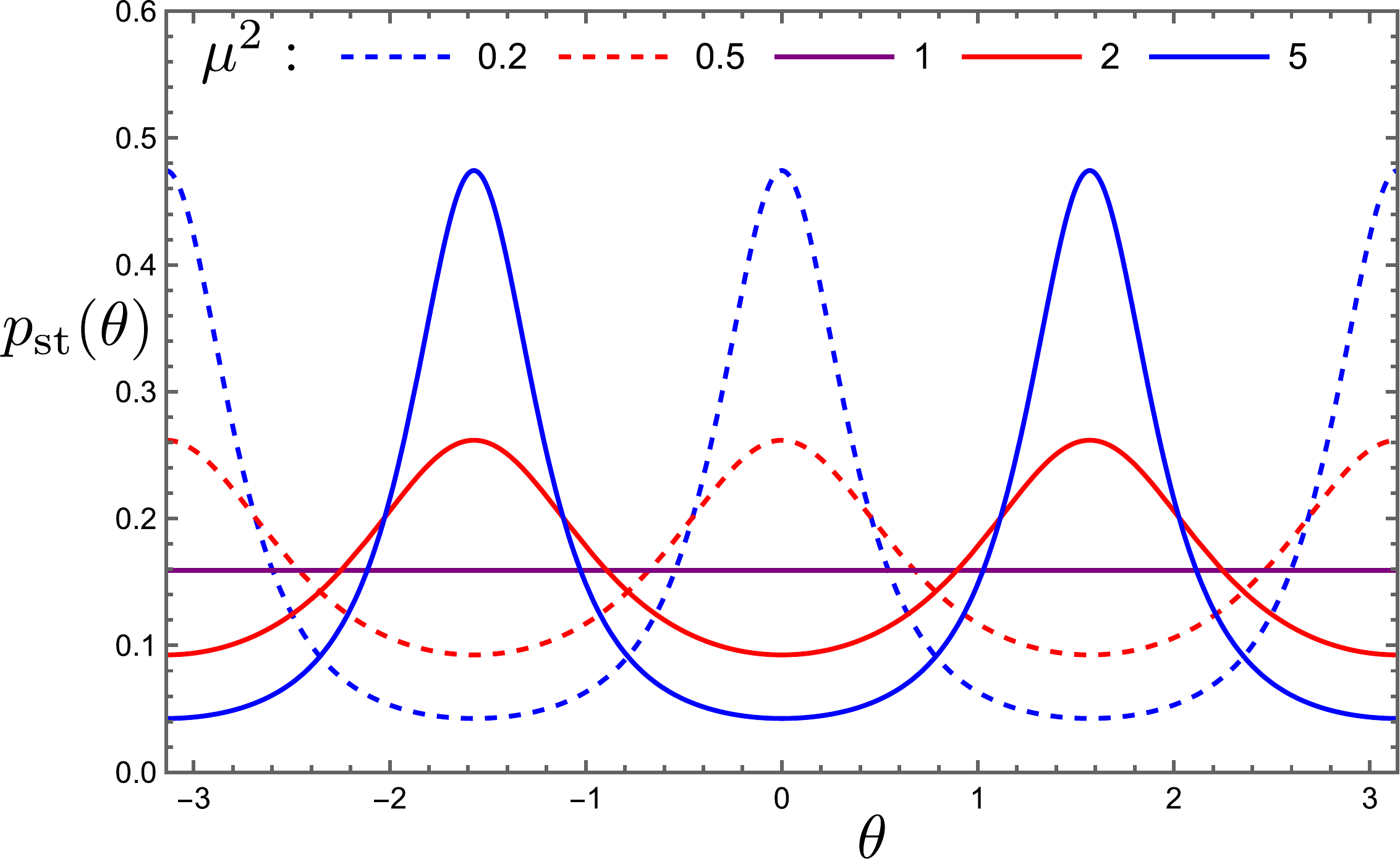}
\par\end{centering}
\caption{Stationary pdfs $p_{{\rm st}}(\theta)$ for simultaneous measurement
of $\sigma_{x}$ and $\sigma_{z}$ with strengths $\alpha_{x}$ and
$\alpha_{z}$, respectively, and strength ratio $\mu=\alpha_{x}/\alpha_{z}$,
when the system is a pure state. \label{fig:stationary-pdfs-over}}

\end{figure}

We therefore return to Eq. (\ref{two SDEs}), set $r_{x}=\sin\theta$,
$r_{z}=\cos\theta$ and derive an SDE for $\theta$ in the form
\begin{align}
d\theta & =\left(\alpha_{x}^{2}-\alpha_{z}^{2}\right)\sin2\theta dt+2\alpha_{x}\cos\theta\,dW_{x}-2\alpha_{z}\sin\theta\,dW_{z}\nonumber \\
 & =\left(\alpha_{x}^{2}-\alpha_{z}^{2}\right)\sin2\theta dt+2\left(\alpha_{x}^{2}\cos^{2}\theta+\alpha_{z}^{2}\sin^{2}\theta\right)^{1/2}dW,\label{dtheta}
\end{align}
which depends on the two measurement strengths $\alpha_{x}$ and $\alpha_{z}$,
and where $dW$ is a Wiener increment. The Fokker-Planck equation
for the pdf $p(\theta,t)$ reads
\begin{align}
\frac{\partial p(\theta,t)}{\partial t} & =-\frac{\partial}{\partial\theta}\Big[\left(\alpha_{x}^{2}-\alpha_{z}^{2}\right)\sin2\theta\,p(\theta,t)\label{FPE theta}\\
 & -2\frac{\partial}{\partial\theta}\left(\alpha_{x}^{2}\cos^{2}\theta+\alpha_{z}^{2}\sin^{2}\theta\right)p(\theta,t)\Big],
\end{align}
and has stationary solutions (with zero probability current) given
by
\begin{equation}
p_{{\rm st}}(\theta)=\frac{\sqrt{2}\mu^{2}\left(1+\mu^{2}-\left(1-\mu^{2}\right)\cos2\theta\right)^{-3/2}}{E\left(1-\mu^{2}\right)+\mu E\left(1-\mu^{-2}\right)},\label{eq:stationary pdf theta}
\end{equation}
where $E(x)=\int_{0}^{\pi/2}\left(1-x\sin^{2}\phi\right)^{1/2}d\phi$
is the complete elliptical integral of the second kind and $\mu=\alpha_{x}/\alpha_{z}$
is the ratio of the two measurement strengths. Examples of stationary
pdfs for various values of $\mu$ are shown in Fig. \ref{fig:stationary-pdfs-over}.
Clearly a greater strength of measurement of observable $\sigma_{x}$
produces higher probability density in the vicinity of the eigenstates
of $\sigma_{x}$ at $\theta=\pm\pi/2$ than in the vicinity of the
eigenstates of $\sigma_{z}$ at $\theta=0$ and $\pi$, and vice versa.

Note that a form of Heisenberg uncertainty is exhibited by the stationary
pdf. In quantum state diffusion, $r_{x}={\rm Tr}(\rho\sigma_{x})$
and $r_{z}={\rm Tr}(\rho\sigma_{z})$ are properties of the quantum
state that are correlated in their evolution. The expectation value
of each in the stationary state is zero:
\begin{align}
\langle r_{z}\rangle & =\int_{-\pi}^{\pi}\cos\theta\,p_{{\rm st}}(\theta)d\theta=0\nonumber \\
\langle r_{x}\rangle & =\int_{-\pi}^{\pi}\sin\theta\,p_{{\rm st}}(\theta)d\theta=0,\label{eq:heisenberg}
\end{align}
while the variances $\langle r_{z}^{2}\rangle-\langle r_{z}\rangle^{2}=\int_{-\pi}^{\pi}\cos^{2}\theta\,p_{{\rm st}}(\theta)d\theta$
and $\langle r_{x}^{2}\rangle-\langle r_{x}\rangle^{2}=\int_{-\pi}^{\pi}\sin^{2}\theta\,p_{{\rm st}}(\theta)d\theta$
sum to unity. A higher measurement strength for one observable drives
up the variance of the associated variable (namely the adopted values
lie close to either 1 or $-1$) while driving down the variance of
the other variable (the value of which lies close to zero). 

\begin{figure}
\begin{centering}
\includegraphics[width=1\columnwidth]{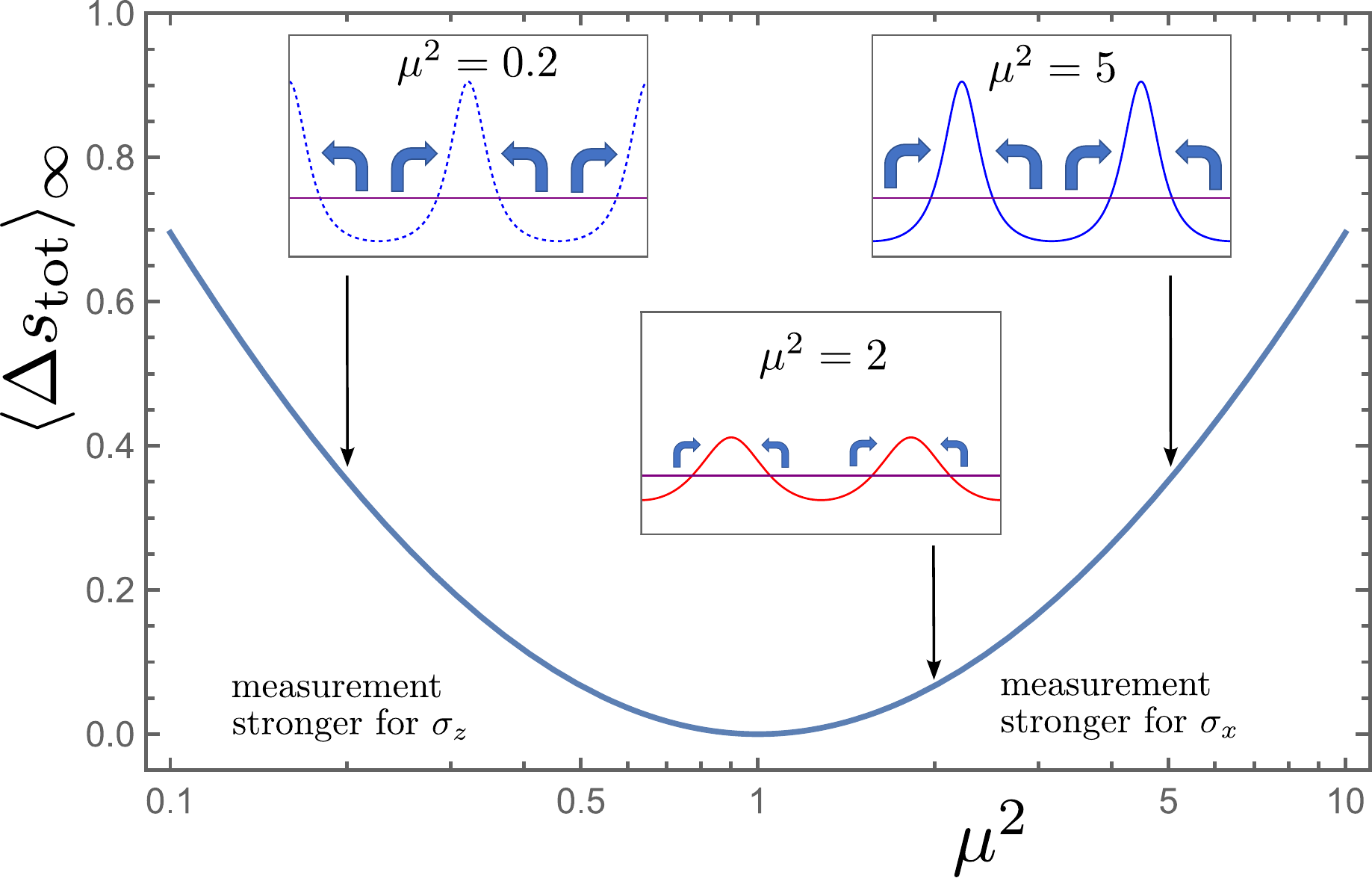}
\par\end{centering}
\caption{The asymptotic mean stochastic entropy production brought about by
an abrupt change in the ratio $\mu=\alpha_{x}/\alpha_{z}$, starting
from equal measurement strengths. The final stationary pdfs for $\mu^{2}=0.2$,
2 and 5, from Fig. \ref{fig:stationary-pdfs-over}, as well as the
initial uniform state, are shown in the insets together with arrows
indicating the change in shape brought about by the process. \label{fig:Mean-dstot-for}}

\end{figure}

The stochastic entropy production associated with the dynamics of
$\theta$ is specified by $A_{\theta}^{{\rm rev}}=0$, $A_{\theta}^{{\rm irr}}=\left(\alpha_{x}^{2}-\alpha_{z}^{2}\right)\sin2\theta$,
and $D_{\theta}=2\left(\alpha_{x}^{2}\cos^{2}\theta+\alpha_{z}^{2}\sin^{2}\theta\right)$
which leads to 
\begin{align}
d\Delta s_{{\rm tot}} & =\left(6\left(\alpha_{x}^{2}-\alpha_{z}^{2}\right)\cos2\theta+\frac{9\left(\alpha_{x}^{2}-\alpha_{z}^{2}\right)^{2}\sin^{2}2\theta}{2\left(\alpha_{x}^{2}\cos^{2}\theta+\alpha_{z}^{2}\sin^{2}\theta\right)}\right)dt\nonumber \\
 & +\frac{3\left(\alpha_{x}^{2}-\alpha_{z}^{2}\right)\sin2\theta}{\left(\alpha_{x}^{2}\cos^{2}\theta+\alpha_{z}^{2}\sin^{2}\theta\right)^{1/2}}dW-d\ln p(\theta,t).\label{eq:dstot theta}
\end{align}
The dynamic and entropic consequences of changing the ratio of measurement
strengths, for an initially pure state, can be established by solving
Eqs. (\ref{dtheta}), (\ref{FPE theta}) and (\ref{eq:dstot theta})
for a given protocol. However, we instead focus attention on a case
with an analytic result. The asymptotic mean production of stochastic
entropy for a transition from a uniform stationary pdf over $\theta$,
at equal measurement strengths $\alpha_{x}^{i}=\alpha_{z}^{i}$, to
a final stationary state brought about by an abrupt change in measurement
strengths to $\alpha_{x}^{f}=\mu\alpha_{z}^{f}$ at $t=0$, takes
the form of a Kullback-Leibler divergence or relative entropy, an
often used measure of distance between probability densities:
\begin{equation}
\langle\Delta s_{{\rm tot}}\rangle_{\infty}=\int p_{{\rm st}}^{i}(\theta)\ln\left(p_{{\rm st}}^{i}(\theta)/p_{{\rm st}}^{f}(\theta)\right)d\theta,\label{eq:mean dstot for theta}
\end{equation}
where the $p_{{\rm st}}^{i,f}(\theta)$ correspond to Eq. (\ref{eq:stationary pdf theta})
with the insertion of $\alpha_{x}^{i,f}$ and $\alpha_{z}^{i,f}$.
This can be derived by noting that $d\Delta s_{{\rm tot}}=-d\ln\left(p(\theta,t)/p_{{\rm st}}(\theta)\right)$
in this case. We plot $\langle\Delta s_{{\rm tot}}\rangle_{\infty}$
for various ratios of final measurement strengths $\mu$ in Fig. \ref{fig:Mean-dstot-for}.
Note that elevation of the measurement strength of one of the observables
relative to the other leads to positive mean stochastic entropy production,
in accordance with the second law, and the effect for enhanced measurement
of $\sigma_{x}$ relative to $\sigma_{z}$ is the same as for enhanced
measurement of $\sigma_{z}$, i.e. the same production emerges for
ratios $\mu$ and $1/\mu$.

\section{Interpretation\label{sec:Interpretation} }

We return now to the physical interpretation of stochastic entropy
production in open quantum systems. By analogy with situations in
classical dynamics, the average of the stochastic entropy production
$\Delta s_{{\rm tot}}$ that accompanies the evolution expresses change
in subjective uncertainty concerning the details of the quantum state
of the world. We have argued that this uncertainty is generated in
the same way as in classical physics. We have taken the dynamical
evolution of the world to be deterministic, but we do not or cannot
attempt to solve the equations of motion for the coordinates exactly.
We instead coarse-grain aspects of the description and employ a set
of stochastic equations that capture the resulting unpredictability
in evolution, again just as in a classical situation. Such modelling
methods can only provide statistical predictions, and hence are characterised
by an increase in entropy of (our perception of) the world. This is
not a physical effect, but merely a measure of the absence of subjective
knowledge, again just as in classical thermodynamics. The key point
is that we take the quantum state vector of the world, and hence the
reduced density matrix of an open system, to be the appropriate physical
description, analogous to classical phase space coordinates.

It is possible to build such stochastic models from an underlying
Hamiltonian describing the system and environment \citep{matos2022},
but here we have adopted a more direct approach, using a framework
of quantum state diffusion to represent the environmental disturbances.
The resulting Markovian stochastic rules of evolution, specified by
Kraus operators, are designed to drive a system continuously and (pseudo)randomly
towards one of its eigenstates. This is our conception of the process
of quantum measurement, in contrast to instantaneous projection. The
resulting evolution of the reduced density matrix resembles a path
taken by a Brownian particle, and it can be described using a Fokker-Planck
equation for a pdf over a suitable phase space, or an It\^o process
that specifies a stochastic trajectory.

The purpose of stochastic entropy production, in both classical and
quantum systems, is to provide a measure of the apparent irreversibility
of evolution and hence an arrow of time. Both of these depend on the
scale of the coarse-graining. The definition in Eq. (\ref{ds_tot})
involves a comparison between the likelihoods, computed according
to the stochastic model employed, of forward and backward sequences
of events. A departure of $\Delta s_{{\rm tot}}$ from zero indicates
that the model dynamics generate one of these sequences preferentially;
that the dynamics are irreversible in the sense of breaking time reversal
symmetry. The preferred sequences will exhibit effects such as dispersion
rather than assembly. 

Nevertheless, parts of the world can become better defined as time
evolves according to these models. Entropy production in a quantum
framework can be used to characterise the approach of an open system
towards an eigenstate under measurement, but also more generally towards
a stationary state in some circumstances. Entropic cost of quantum
measurement is analogous to such a cost in simple models of classical
measurement \citep{Ford16}. Furthermore, we can conceive of quantum
processes that are reversible, in the sense that the average of $\Delta s_{{\rm tot}}$
is zero. This would arise, as in classical circumstances, when the
driving of the system, for example the rate of change of coupling
to a measuring device, becomes quasistatic. Hence quantum measurement
need not be irreversible, neither in the dynamic nor in the entropic
sense.

\section{Conclusions\label{sec:Conclusions}}

Entropy production represents increasing subjective uncertainty of
microscopic configuration brought about by employing stochastic models
of the dynamics instead of the underlying deterministic equations
of motion that are responsible for complex, dispersive behaviour.
These ideas can apply to quantum systems, for which we regard the
reduced density matrix as a physical property analogous to a set of
physical coordinates of a classical system. The reduced density matrix
evolves pseudorandomly through interactions with an underspecified
environment, which we represent in a minimal fashion using Kraus operators
and a framework of Markovian quantum state diffusion. We concern ourselves
with the uncertainty in the reduced density matrix that is actually
adopted by the system. Stochastic entropy production can then be computed
using analysis of the relative probabilities of forward and backward
Brownian trajectories of the reduced density matrix.

The crucial features of quantum mechanics are captured by such a dynamics,
in particular the stochastic selection of an eigenstate according
to the Born rule. A further feature has been explored, for a simple
two level system, where the simultaneous measurement of two observables
represented by non-commuting operators can be considered. The system
is prevented from selecting an eigenstate of either operator, in line
with expectation, and instead adopts a state of correlated stationary
uncertainty with respect to the two observables.

The model of measurement used here has the effect of purifying the
system, i.e. eliminating any initial entanglement between the system
and its environment. The effect is a consequence of the simplicity
of the model, but it is perfectly in line with the idea that a system
takes an eigenstate of a system observable after the process of measurement.
The final state of the environment (the measuring device) is nevertheless
correlated with the final state of the system, and this is the means
by which it is able to convey information about the system observable
and preserve a record of the measurement.

We suggest that the reduced density matrix typically used to describe
an open quantum system is an average over an ensemble of adoptable
states; pure as well as those entangled with the environment. Moreover,
the ensemble average is not suitable for modelling eigenstate selection,
which takes place at the level of ensemble members. This problem is
traditionally accommodated by introducing a process of projective
measurement that takes place outside the regular dynamics and changes
the ensemble average, but such a difficulty is not present when considering
the dynamics of ensemble members.

The dynamics we employ therefore conceptualise quantum mechanics as
the evolution of physical properties that behave in a complex but
relatively unmysterious fashion. The quantum state is more than a
provider of information about probabilities of projective measurement
outcomes. The reduced density matrix, and by implication the quantum
state vector of the world, are treated as physical coordinates and
not merely bearers of information.

Using such a dynamical framework, the main purpose of this paper has
been to provide explicit examples of stochastic entropy production
for a simple open quantum system, and to suggest that this quantity
is the most appropriate extension into the quantum regime of the modern
concept of entropy production. We have studied stochastic entropy
production for scenarios involving the measurement of one and then
two observables. Mean stochastic entropy production in this context
measures the change in subjective uncertainty concerning the adopted
quantum state of the world. It never decreases, thus satisfying the
second law of thermodynamics. The von Neumann entropy is a measure
of uncertainty in measurement outcome, but compared to mean stochastic
entropy production it plays a rather different role. The connections
between the two are worth exploring further.
\begin{acknowledgments}
This work was supported by the U.K. Engineering and Physical Sciences
Research Council through the Centre for Doctoral Training in Delivering
Quantum Technologies at UCL, grant number 1489394.
\end{acknowledgments}

\bibliography{Uncertain_density_matrix}

\begin{thebibliography}{50}%
\makeatletter
\providecommand \@ifxundefined [1]{%
 \@ifx{#1\undefined}
}%
\providecommand \@ifnum [1]{%
 \ifnum #1\expandafter \@firstoftwo
 \else \expandafter \@secondoftwo
 \fi
}%
\providecommand \@ifx [1]{%
 \ifx #1\expandafter \@firstoftwo
 \else \expandafter \@secondoftwo
 \fi
}%
\providecommand \natexlab [1]{#1}%
\providecommand \enquote  [1]{``#1''}%
\providecommand \bibnamefont  [1]{#1}%
\providecommand \bibfnamefont [1]{#1}%
\providecommand \citenamefont [1]{#1}%
\providecommand \href@noop [0]{\@secondoftwo}%
\providecommand \href [0]{\begingroup \@sanitize@url \@href}%
\providecommand \@href[1]{\@@startlink{#1}\@@href}%
\providecommand \@@href[1]{\endgroup#1\@@endlink}%
\providecommand \@sanitize@url [0]{\catcode `\\12\catcode `\$12\catcode
  `\&12\catcode `\#12\catcode `\^12\catcode `\_12\catcode `\%12\relax}%
\providecommand \@@startlink[1]{}%
\providecommand \@@endlink[0]{}%
\providecommand \url  [0]{\begingroup\@sanitize@url \@url }%
\providecommand \@url [1]{\endgroup\@href {#1}{\urlprefix }}%
\providecommand \urlprefix  [0]{URL }%
\providecommand \Eprint [0]{\href }%
\providecommand \doibase [0]{https://doi.org/}%
\providecommand \selectlanguage [0]{\@gobble}%
\providecommand \bibinfo  [0]{\@secondoftwo}%
\providecommand \bibfield  [0]{\@secondoftwo}%
\providecommand \translation [1]{[#1]}%
\providecommand \BibitemOpen [0]{}%
\providecommand \bibitemStop [0]{}%
\providecommand \bibitemNoStop [0]{.\EOS\space}%
\providecommand \EOS [0]{\spacefactor3000\relax}%
\providecommand \BibitemShut  [1]{\csname bibitem#1\endcsname}%
\let\auto@bib@innerbib\@empty
\bibitem [{\citenamefont {Lebowitz}(1993)}]{lebowitz1993boltzmann}%
  \BibitemOpen
  \bibfield  {author} {\bibinfo {author} {\bibfnamefont {J.~L.}\ \bibnamefont
  {Lebowitz}},\ }\bibfield  {title} {\bibinfo {title} {Boltzmann's entropy and
  time's arrow},\ }\href@noop {} {\bibfield  {journal} {\bibinfo  {journal}
  {Physics Today}\ }\textbf {\bibinfo {volume} {46}},\ \bibinfo {pages} {32}
  (\bibinfo {year} {1993})}\BibitemShut {NoStop}%
\bibitem [{\citenamefont {Albert}(2009)}]{albert2009time}%
  \BibitemOpen
  \bibfield  {author} {\bibinfo {author} {\bibfnamefont {D.~Z.}\ \bibnamefont
  {Albert}},\ }\href@noop {} {\emph {\bibinfo {title} {Time and Chance}}}\
  (\bibinfo  {publisher} {Harvard University Press},\ \bibinfo {year}
  {2009})\BibitemShut {NoStop}%
\bibitem [{\citenamefont {{I. J. Ford}}(2013)}]{Ford-book2013}%
  \BibitemOpen
  \bibfield  {author} {\bibinfo {author} {\bibnamefont {{I. J. Ford}}},\
  }\href@noop {} {\emph {\bibinfo {title} {Statistical {Physics}: an {Entropic}
  {Approach}}}}\ (\bibinfo  {publisher} {Wiley},\ \bibinfo {year}
  {2013})\BibitemShut {NoStop}%
\bibitem [{\citenamefont {Seifert}(2008)}]{seifert2008stochastic}%
  \BibitemOpen
  \bibfield  {author} {\bibinfo {author} {\bibfnamefont {U.}~\bibnamefont
  {Seifert}},\ }\bibfield  {title} {\bibinfo {title} {Stochastic
  thermodynamics: principles and perspectives},\ }\href@noop {} {\bibfield
  {journal} {\bibinfo  {journal} {The European Physical Journal B}\ }\textbf
  {\bibinfo {volume} {64}},\ \bibinfo {pages} {423} (\bibinfo {year}
  {2008})}\BibitemShut {NoStop}%
\bibitem [{\citenamefont {Harris}\ and\ \citenamefont
  {Sch{\"u}tz}(2007)}]{harris2007fluctuation}%
  \BibitemOpen
  \bibfield  {author} {\bibinfo {author} {\bibfnamefont {R.~J.}\ \bibnamefont
  {Harris}}\ and\ \bibinfo {author} {\bibfnamefont {G.~M.}\ \bibnamefont
  {Sch{\"u}tz}},\ }\bibfield  {title} {\bibinfo {title} {Fluctuation theorems
  for stochastic dynamics},\ }\href@noop {} {\bibfield  {journal} {\bibinfo
  {journal} {Journal of Statistical Mechanics: Theory and Experiment}\ }\textbf
  {\bibinfo {volume} {2007}},\ \bibinfo {pages} {P07020} (\bibinfo {year}
  {2007})}\BibitemShut {NoStop}%
\bibitem [{\citenamefont {{R. E. Spinney and I. J.
  Ford}}(2012)}]{SpinneyFord12a}%
  \BibitemOpen
  \bibfield  {author} {\bibinfo {author} {\bibnamefont {{R. E. Spinney and I.
  J. Ford}}},\ }\bibfield  {title} {\bibinfo {title} {Nonequilibrium
  thermodynamics of stochastic systems with odd and even variables},\
  }\href@noop {} {\bibfield  {journal} {\bibinfo  {journal} {Physical Review
  Letters}\ }\textbf {\bibinfo {volume} {108}},\ \bibinfo {pages} {170603}
  (\bibinfo {year} {2012})}\BibitemShut {NoStop}%
\bibitem [{\citenamefont {Spinney}\ and\ \citenamefont
  {Ford}(2012)}]{spinney2012entropy}%
  \BibitemOpen
  \bibfield  {author} {\bibinfo {author} {\bibfnamefont {R.~E.}\ \bibnamefont
  {Spinney}}\ and\ \bibinfo {author} {\bibfnamefont {I.~J.}\ \bibnamefont
  {Ford}},\ }\bibfield  {title} {\bibinfo {title} {Entropy production in full
  phase space for continuous stochastic dynamics},\ }\href@noop {} {\bibfield
  {journal} {\bibinfo  {journal} {Physical Review E}\ }\textbf {\bibinfo
  {volume} {85}},\ \bibinfo {pages} {051113} (\bibinfo {year}
  {2012})}\BibitemShut {NoStop}%
\bibitem [{\citenamefont {Ford}\ \emph {et~al.}(2015)\citenamefont {Ford},
  \citenamefont {Laker},\ and\ \citenamefont
  {Charlesworth}}]{ford2015stochastic}%
  \BibitemOpen
  \bibfield  {author} {\bibinfo {author} {\bibfnamefont {I.~J.}\ \bibnamefont
  {Ford}}, \bibinfo {author} {\bibfnamefont {Z.~P.}\ \bibnamefont {Laker}},\
  and\ \bibinfo {author} {\bibfnamefont {H.~J.}\ \bibnamefont {Charlesworth}},\
  }\bibfield  {title} {\bibinfo {title} {Stochastic entropy production arising
  from nonstationary thermal transport},\ }\href@noop {} {\bibfield  {journal}
  {\bibinfo  {journal} {Physical Review E}\ }\textbf {\bibinfo {volume} {92}},\
  \bibinfo {pages} {042108} (\bibinfo {year} {2015})}\BibitemShut {NoStop}%
\bibitem [{\citenamefont {Wiseman}(1996)}]{wiseman1996}%
  \BibitemOpen
  \bibfield  {author} {\bibinfo {author} {\bibfnamefont {H.~M.}\ \bibnamefont
  {Wiseman}},\ }\bibfield  {title} {\bibinfo {title} {Quantum trajectories and
  quantum measurement theory},\ }\href
  {https://doi.org/10.1088/1355-5111/8/1/015} {\bibfield  {journal} {\bibinfo
  {journal} {Quantum and Semiclassical Optics}\ }\textbf {\bibinfo {volume}
  {8}},\ \bibinfo {pages} {205} (\bibinfo {year} {1996})}\BibitemShut {NoStop}%
\bibitem [{\citenamefont {Breuer}\ and\ \citenamefont
  {Petruccione}(2007)}]{breuer2007}%
  \BibitemOpen
  \bibfield  {author} {\bibinfo {author} {\bibfnamefont {H.-P.}\ \bibnamefont
  {Breuer}}\ and\ \bibinfo {author} {\bibfnamefont {F.}~\bibnamefont
  {Petruccione}},\ }\href
  {https://doi.org/10.1093/acprof:oso/9780199213900.001.0001} {\emph {\bibinfo
  {title} {The {{Theory}} of {{Open Quantum Systems}}}}}\ (\bibinfo
  {publisher} {{Oxford University Press}},\ \bibinfo {year} {2007})\BibitemShut
  {NoStop}%
\bibitem [{\citenamefont {Weiss}(2012)}]{weiss2012quantum}%
  \BibitemOpen
  \bibfield  {author} {\bibinfo {author} {\bibfnamefont {U.}~\bibnamefont
  {Weiss}},\ }\href@noop {} {\emph {\bibinfo {title} {Quantum {Dissipative}
  {Systems}}}}\ (\bibinfo  {publisher} {World {Scientific}},\ \bibinfo {year}
  {2012})\BibitemShut {NoStop}%
\bibitem [{\citenamefont {Brun}(2000)}]{brun2000continuous}%
  \BibitemOpen
  \bibfield  {author} {\bibinfo {author} {\bibfnamefont {T.~A.}\ \bibnamefont
  {Brun}},\ }\bibfield  {title} {\bibinfo {title} {Continuous measurements,
  quantum trajectories, and decoherent histories},\ }\href@noop {} {\bibfield
  {journal} {\bibinfo  {journal} {Physical Review A}\ }\textbf {\bibinfo
  {volume} {61}},\ \bibinfo {pages} {042107} (\bibinfo {year}
  {2000})}\BibitemShut {NoStop}%
\bibitem [{\citenamefont {Jacobs}(2014)}]{jacobs2014quantum}%
  \BibitemOpen
  \bibfield  {author} {\bibinfo {author} {\bibfnamefont {K.}~\bibnamefont
  {Jacobs}},\ }\href@noop {} {\emph {\bibinfo {title} {Quantum {Measurement}
  {Theory} and its {Applications}}}}\ (\bibinfo  {publisher} {Cambridge
  University Press},\ \bibinfo {year} {2014})\BibitemShut {NoStop}%
\bibitem [{\citenamefont {Jacobs}\ and\ \citenamefont
  {Steck}(2006)}]{jacobs2006straightforward}%
  \BibitemOpen
  \bibfield  {author} {\bibinfo {author} {\bibfnamefont {K.}~\bibnamefont
  {Jacobs}}\ and\ \bibinfo {author} {\bibfnamefont {D.~A.}\ \bibnamefont
  {Steck}},\ }\bibfield  {title} {\bibinfo {title} {A straightforward
  introduction to continuous quantum measurement},\ }\href@noop {} {\bibfield
  {journal} {\bibinfo  {journal} {Contemporary Physics}\ }\textbf {\bibinfo
  {volume} {47}},\ \bibinfo {pages} {279} (\bibinfo {year} {2006})}\BibitemShut
  {NoStop}%
\bibitem [{\citenamefont {Srednicki}(1994)}]{srednicki1994chaos}%
  \BibitemOpen
  \bibfield  {author} {\bibinfo {author} {\bibfnamefont {M.}~\bibnamefont
  {Srednicki}},\ }\bibfield  {title} {\bibinfo {title} {Chaos and quantum
  thermalization},\ }\href@noop {} {\bibfield  {journal} {\bibinfo  {journal}
  {Physical Review E}\ }\textbf {\bibinfo {volume} {50}},\ \bibinfo {pages}
  {888} (\bibinfo {year} {1994})}\BibitemShut {NoStop}%
\bibitem [{\citenamefont {Percival}(1998)}]{percival1998}%
  \BibitemOpen
  \bibfield  {author} {\bibinfo {author} {\bibfnamefont {I.}~\bibnamefont
  {Percival}},\ }\href@noop {} {\emph {\bibinfo {title} {Quantum {{State
  Diffusion}}}}}\ (\bibinfo  {publisher} {{Cambridge University Press}},\
  \bibinfo {year} {1998})\BibitemShut {NoStop}%
\bibitem [{\citenamefont {Strunz}(1996)}]{strunz1996linear}%
  \BibitemOpen
  \bibfield  {author} {\bibinfo {author} {\bibfnamefont {W.~T.}\ \bibnamefont
  {Strunz}},\ }\bibfield  {title} {\bibinfo {title} {Linear quantum state
  diffusion for non-markovian open quantum systems},\ }\href@noop {} {\bibfield
   {journal} {\bibinfo  {journal} {Physics Letters A}\ }\textbf {\bibinfo
  {volume} {224}},\ \bibinfo {pages} {25} (\bibinfo {year} {1996})}\BibitemShut
  {NoStop}%
\bibitem [{\citenamefont {Gisin}\ and\ \citenamefont
  {Percival}(1992)}]{gisin1992quantum}%
  \BibitemOpen
  \bibfield  {author} {\bibinfo {author} {\bibfnamefont {N.}~\bibnamefont
  {Gisin}}\ and\ \bibinfo {author} {\bibfnamefont {I.~C.}\ \bibnamefont
  {Percival}},\ }\bibfield  {title} {\bibinfo {title} {The quantum-state
  diffusion model applied to open systems},\ }\href@noop {} {\bibfield
  {journal} {\bibinfo  {journal} {Journal of Physics A: Mathematical and
  General}\ }\textbf {\bibinfo {volume} {25}},\ \bibinfo {pages} {5677}
  (\bibinfo {year} {1992})}\BibitemShut {NoStop}%
\bibitem [{\citenamefont {Gisin}\ and\ \citenamefont
  {Percival}(1993)}]{gisin1993quantum}%
  \BibitemOpen
  \bibfield  {author} {\bibinfo {author} {\bibfnamefont {N.}~\bibnamefont
  {Gisin}}\ and\ \bibinfo {author} {\bibfnamefont {I.~C.}\ \bibnamefont
  {Percival}},\ }\bibfield  {title} {\bibinfo {title} {Quantum state diffusion,
  localization and quantum dispersion entropy},\ }\href@noop {} {\bibfield
  {journal} {\bibinfo  {journal} {Journal of Physics A: Mathematical and
  General}\ }\textbf {\bibinfo {volume} {26}},\ \bibinfo {pages} {2233}
  (\bibinfo {year} {1993})}\BibitemShut {NoStop}%
\bibitem [{\citenamefont {Strunz}\ \emph {et~al.}(1999)\citenamefont {Strunz},
  \citenamefont {Di{\'o}si},\ and\ \citenamefont {Gisin}}]{strunz1999open}%
  \BibitemOpen
  \bibfield  {author} {\bibinfo {author} {\bibfnamefont {W.~T.}\ \bibnamefont
  {Strunz}}, \bibinfo {author} {\bibfnamefont {L.}~\bibnamefont {Di{\'o}si}},\
  and\ \bibinfo {author} {\bibfnamefont {N.}~\bibnamefont {Gisin}},\ }\bibfield
   {title} {\bibinfo {title} {Open system dynamics with non-{Markovian} quantum
  trajectories},\ }\href@noop {} {\bibfield  {journal} {\bibinfo  {journal}
  {Physical Review Letters}\ }\textbf {\bibinfo {volume} {82}},\ \bibinfo
  {pages} {1801} (\bibinfo {year} {1999})}\BibitemShut {NoStop}%
\bibitem [{\citenamefont {Jordan}(2013)}]{jordan2013}%
  \BibitemOpen
  \bibfield  {author} {\bibinfo {author} {\bibfnamefont {A.~N.}\ \bibnamefont
  {Jordan}},\ }\bibfield  {title} {\bibinfo {title} {Watching the wavefunction
  collapse},\ }\href {https://doi.org/10.1038/502177a} {\bibfield  {journal}
  {\bibinfo  {journal} {Nature}\ }\textbf {\bibinfo {volume} {502}},\ \bibinfo
  {pages} {177} (\bibinfo {year} {2013})}\BibitemShut {NoStop}%
\bibitem [{\citenamefont {Vinjanampathy}\ and\ \citenamefont
  {Anders}(2016)}]{Vinjamampathy16}%
  \BibitemOpen
  \bibfield  {author} {\bibinfo {author} {\bibfnamefont {S.}~\bibnamefont
  {Vinjanampathy}}\ and\ \bibinfo {author} {\bibfnamefont {J.}~\bibnamefont
  {Anders}},\ }\bibfield  {title} {\bibinfo {title} {Quantum thermodynamics},\
  }\href@noop {} {\bibfield  {journal} {\bibinfo  {journal} {Contemporary
  Physics}\ }\textbf {\bibinfo {volume} {57}},\ \bibinfo {pages} {545}
  (\bibinfo {year} {2016})}\BibitemShut {NoStop}%
\bibitem [{\citenamefont {Kammerlander}\ and\ \citenamefont
  {Anders}(2016)}]{kammerlander2016coherence}%
  \BibitemOpen
  \bibfield  {author} {\bibinfo {author} {\bibfnamefont {P.}~\bibnamefont
  {Kammerlander}}\ and\ \bibinfo {author} {\bibfnamefont {J.}~\bibnamefont
  {Anders}},\ }\bibfield  {title} {\bibinfo {title} {Coherence and measurement
  in quantum thermodynamics},\ }\href@noop {} {\bibfield  {journal} {\bibinfo
  {journal} {Scientific Reports}\ }\textbf {\bibinfo {volume} {6}} (\bibinfo
  {year} {2016})}\BibitemShut {NoStop}%
\bibitem [{\citenamefont {Walls}\ \emph {et~al.}(2024)\citenamefont {Walls},
  \citenamefont {Schachter}, \citenamefont {{H. Qian}},\ and\ \citenamefont
  {Ford}}]{Walls24}%
  \BibitemOpen
  \bibfield  {author} {\bibinfo {author} {\bibfnamefont {S.~M.}\ \bibnamefont
  {Walls}}, \bibinfo {author} {\bibfnamefont {J.~M.}\ \bibnamefont
  {Schachter}}, \bibinfo {author} {\bibnamefont {{H. Qian}}},\ and\ \bibinfo
  {author} {\bibfnamefont {I.~J.}\ \bibnamefont {Ford}},\ }\bibfield  {title}
  {\bibinfo {title} {Stochastic quantum trajectories demonstrate the {Quantum
  Zeno Effect} in open spin 1/2, spin 1 and spin 3/2 systems},\ }\href@noop {}
  {\bibfield  {journal} {\bibinfo  {journal} {J. Phys. A: Math. Theor.}\
  }\textbf {\bibinfo {volume} {57}},\ \bibinfo {pages} {175301} (\bibinfo
  {year} {2024})}\BibitemShut {NoStop}%
\bibitem [{\citenamefont {Minev}\ \emph {et~al.}(2019)\citenamefont {Minev},
  \citenamefont {Mundhada}, \citenamefont {Shankar}, \citenamefont {Reinhold},
  \citenamefont {Gutiérrez-Jáuregui}, \citenamefont {Schoelkopf}, \citenamefont
  {Mirrahimi}, \citenamefont {Carmichael},\ and\ \citenamefont
  {Devoret}}]{Minev19}%
  \BibitemOpen
  \bibfield  {author} {\bibinfo {author} {\bibfnamefont {Z.}~\bibnamefont
  {Minev}}, \bibinfo {author} {\bibfnamefont {S.}~\bibnamefont {Mundhada}},
  \bibinfo {author} {\bibfnamefont {S.}~\bibnamefont {Shankar}}, \bibinfo
  {author} {\bibfnamefont {P.}~\bibnamefont {Reinhold}}, \bibinfo {author}
  {\bibfnamefont {R.}~\bibnamefont {Gutiérrez-Jáuregui}}, \bibinfo {author}
  {\bibfnamefont {R.}~\bibnamefont {Schoelkopf}}, \bibinfo {author}
  {\bibfnamefont {M.}~\bibnamefont {Mirrahimi}}, \bibinfo {author}
  {\bibfnamefont {H.}~\bibnamefont {Carmichael}},\ and\ \bibinfo {author}
  {\bibfnamefont {M.}~\bibnamefont {Devoret}},\ }\bibfield  {title} {\bibinfo
  {title} {To catch and reverse a quantum jump mid-flight},\ }\href@noop {}
  {\bibfield  {journal} {\bibinfo  {journal} {Nature}\ }\textbf {\bibinfo
  {volume} {570}},\ \bibinfo {pages} {200} (\bibinfo {year}
  {2019})}\BibitemShut {NoStop}%
\bibitem [{\citenamefont {Holland}(2005)}]{holland2005}%
  \BibitemOpen
  \bibfield  {author} {\bibinfo {author} {\bibfnamefont {P.}~\bibnamefont
  {Holland}},\ }\bibfield  {title} {\bibinfo {title} {What's wrong with
  {{Einstein}}'s 1927 hidden-variable interpretation of quantum mechanics?},\
  }\href {https://doi.org/10.1007/s10701-004-1940-7} {\bibfield  {journal}
  {\bibinfo  {journal} {Foundations of Physics}\ }\textbf {\bibinfo {volume}
  {35}},\ \bibinfo {pages} {177} (\bibinfo {year} {2005})}\BibitemShut
  {NoStop}%
\bibitem [{\citenamefont {Wiseman}\ and\ \citenamefont
  {Gambetta}(2008)}]{WiseGamb08}%
  \BibitemOpen
  \bibfield  {author} {\bibinfo {author} {\bibfnamefont {H.~M.}\ \bibnamefont
  {Wiseman}}\ and\ \bibinfo {author} {\bibfnamefont {J.~M.}\ \bibnamefont
  {Gambetta}},\ }\bibfield  {title} {\bibinfo {title} {{Pure-state quantum
  trajectories for general non-Markovian systems do not exist}},\ }\href
  {https://doi.org/10.1103/PhysRevLett.101.140401} {\bibfield  {journal}
  {\bibinfo  {journal} {Physical Review Letters}\ }\textbf {\bibinfo {volume}
  {101}},\ \bibinfo {pages} {140401} (\bibinfo {year} {2008})}\BibitemShut
  {NoStop}%
\bibitem [{\citenamefont {Hiley}\ \emph {et~al.}(2019)\citenamefont {Hiley},
  \citenamefont {{M. A. de Gosson}},\ and\ \citenamefont {Dennis}}]{hiley2019}%
  \BibitemOpen
  \bibfield  {author} {\bibinfo {author} {\bibfnamefont {B.~J.}\ \bibnamefont
  {Hiley}}, \bibinfo {author} {\bibnamefont {{M. A. de Gosson}}},\ and\
  \bibinfo {author} {\bibfnamefont {G.}~\bibnamefont {Dennis}},\ }\bibfield
  {title} {\bibinfo {title} {Quantum {Trajectories}: {Dirac}, {Moyal} and
  {Bohm}},\ }\href {https://doi.org/10.12743/quanta.v8i1.84} {\bibfield
  {journal} {\bibinfo  {journal} {Quanta}\ }\textbf {\bibinfo {volume} {8}},\
  \bibinfo {pages} {11} (\bibinfo {year} {2019})}\BibitemShut {NoStop}%
\bibitem [{\citenamefont {Roch}\ \emph {et~al.}(2014)\citenamefont {Roch},
  \citenamefont {Schwartz}, \citenamefont {Motzoi}, \citenamefont {Macklin},
  \citenamefont {Vijay}, \citenamefont {Eddins}, \citenamefont {Korotkov},
  \citenamefont {Whaley}, \citenamefont {Sarovar},\ and\ \citenamefont
  {Siddiqi}}]{roch2014}%
  \BibitemOpen
  \bibfield  {author} {\bibinfo {author} {\bibfnamefont {N.}~\bibnamefont
  {Roch}}, \bibinfo {author} {\bibfnamefont {M.~E.}\ \bibnamefont {Schwartz}},
  \bibinfo {author} {\bibfnamefont {F.}~\bibnamefont {Motzoi}}, \bibinfo
  {author} {\bibfnamefont {C.}~\bibnamefont {Macklin}}, \bibinfo {author}
  {\bibfnamefont {R.}~\bibnamefont {Vijay}}, \bibinfo {author} {\bibfnamefont
  {A.~W.}\ \bibnamefont {Eddins}}, \bibinfo {author} {\bibfnamefont {A.~N.}\
  \bibnamefont {Korotkov}}, \bibinfo {author} {\bibfnamefont {K.~B.}\
  \bibnamefont {Whaley}}, \bibinfo {author} {\bibfnamefont {M.}~\bibnamefont
  {Sarovar}},\ and\ \bibinfo {author} {\bibfnamefont {I.}~\bibnamefont
  {Siddiqi}},\ }\bibfield  {title} {\bibinfo {title} {Observation of
  {{measurement-induced entanglement}} and {{quantum trajectories}} of {{remote
  superconducting qubits}}},\ }\href
  {https://doi.org/10.1103/PhysRevLett.112.170501} {\bibfield  {journal}
  {\bibinfo  {journal} {Physical Review Letters}\ }\textbf {\bibinfo {volume}
  {112}},\ \bibinfo {pages} {170501} (\bibinfo {year} {2014})}\BibitemShut
  {NoStop}%
\bibitem [{\citenamefont {Gröblacher}\ \emph {et~al.}(2007)\citenamefont
  {Gröblacher}, \citenamefont {Paterek}, \citenamefont {Kaltenbaek},
  \citenamefont {Brukner}, \citenamefont {Zukowski}, \citenamefont
  {Aspelmeyer},\ and\ \citenamefont {Zeilinger}}]{Groeblacher07}%
  \BibitemOpen
  \bibfield  {author} {\bibinfo {author} {\bibfnamefont {S.}~\bibnamefont
  {Gröblacher}}, \bibinfo {author} {\bibfnamefont {T.}~\bibnamefont {Paterek}},
  \bibinfo {author} {\bibfnamefont {R.}~\bibnamefont {Kaltenbaek}}, \bibinfo
  {author} {\bibfnamefont {C.}~\bibnamefont {Brukner}}, \bibinfo {author}
  {\bibfnamefont {M.}~\bibnamefont {Zukowski}}, \bibinfo {author}
  {\bibfnamefont {M.}~\bibnamefont {Aspelmeyer}},\ and\ \bibinfo {author}
  {\bibfnamefont {A.}~\bibnamefont {Zeilinger}},\ }\bibfield  {title} {\bibinfo
  {title} {An experimental test of non-local realism},\ }\href@noop {}
  {\bibfield  {journal} {\bibinfo  {journal} {Nature}\ }\textbf {\bibinfo
  {volume} {446}},\ \bibinfo {pages} {871} (\bibinfo {year}
  {2007})}\BibitemShut {NoStop}%
\bibitem [{\citenamefont {Norsen}(2017)}]{norsen2017a}%
  \BibitemOpen
  \bibfield  {author} {\bibinfo {author} {\bibfnamefont {T.}~\bibnamefont
  {Norsen}},\ }\href@noop {} {\emph {\bibinfo {title} {Foundations of {{Quantum
  Mechanics}}: {{An Exploration}} of the {{Physical Meaning}} of {{Quantum
  Theory}}}}}\ (\bibinfo  {publisher} {{Springer}},\ \bibinfo {year}
  {2017})\BibitemShut {NoStop}%
\bibitem [{\citenamefont {Hossenfelder}\ and\ \citenamefont
  {Palmer}(2020)}]{Hossenfelder20}%
  \BibitemOpen
  \bibfield  {author} {\bibinfo {author} {\bibfnamefont {S.}~\bibnamefont
  {Hossenfelder}}\ and\ \bibinfo {author} {\bibfnamefont {T.}~\bibnamefont
  {Palmer}},\ }\bibfield  {title} {\bibinfo {title} {Rethinking
  superdeterminism},\ }\href@noop {} {\bibfield  {journal} {\bibinfo  {journal}
  {Frontiers in Physics}\ }\textbf {\bibinfo {volume} {8}},\ \bibinfo {pages}
  {139} (\bibinfo {year} {2020})}\BibitemShut {NoStop}%
\bibitem [{\citenamefont {Deffner}\ and\ \citenamefont
  {Lutz}(2011)}]{deffner2011nonequilibrium}%
  \BibitemOpen
  \bibfield  {author} {\bibinfo {author} {\bibfnamefont {S.}~\bibnamefont
  {Deffner}}\ and\ \bibinfo {author} {\bibfnamefont {E.}~\bibnamefont {Lutz}},\
  }\bibfield  {title} {\bibinfo {title} {Nonequilibrium entropy production for
  open quantum systems},\ }\href@noop {} {\bibfield  {journal} {\bibinfo
  {journal} {Physical Review Letters}\ }\textbf {\bibinfo {volume} {107}},\
  \bibinfo {pages} {140404} (\bibinfo {year} {2011})}\BibitemShut {NoStop}%
\bibitem [{\citenamefont {Leggio}\ \emph {et~al.}(2013)\citenamefont {Leggio},
  \citenamefont {Napoli}, \citenamefont {Messina},\ and\ \citenamefont
  {Breuer}}]{leggio2013entropy}%
  \BibitemOpen
  \bibfield  {author} {\bibinfo {author} {\bibfnamefont {B.}~\bibnamefont
  {Leggio}}, \bibinfo {author} {\bibfnamefont {A.}~\bibnamefont {Napoli}},
  \bibinfo {author} {\bibfnamefont {A.}~\bibnamefont {Messina}},\ and\ \bibinfo
  {author} {\bibfnamefont {H.-P.}\ \bibnamefont {Breuer}},\ }\bibfield  {title}
  {\bibinfo {title} {Entropy production and information fluctuations along
  quantum trajectories},\ }\href@noop {} {\bibfield  {journal} {\bibinfo
  {journal} {Physical Review A}\ }\textbf {\bibinfo {volume} {88}},\ \bibinfo
  {pages} {042111} (\bibinfo {year} {2013})}\BibitemShut {NoStop}%
\bibitem [{\citenamefont {Horowitz}\ and\ \citenamefont
  {Parrondo}(2013)}]{horowitz2013entropy}%
  \BibitemOpen
  \bibfield  {author} {\bibinfo {author} {\bibfnamefont {J.~M.}\ \bibnamefont
  {Horowitz}}\ and\ \bibinfo {author} {\bibfnamefont {J.~M.}\ \bibnamefont
  {Parrondo}},\ }\bibfield  {title} {\bibinfo {title} {Entropy production along
  nonequilibrium quantum jump trajectories},\ }\href@noop {} {\bibfield
  {journal} {\bibinfo  {journal} {New Journal of Physics}\ }\textbf {\bibinfo
  {volume} {15}},\ \bibinfo {pages} {085028} (\bibinfo {year}
  {2013})}\BibitemShut {NoStop}%
\bibitem [{\citenamefont {Elouard}\ \emph
  {et~al.}(2017{\natexlab{a}})\citenamefont {Elouard}, \citenamefont
  {Bernardes}, \citenamefont {Carvalho}, \citenamefont {Santos},\ and\
  \citenamefont {Auff{\`e}ves}}]{elouard2017probing}%
  \BibitemOpen
  \bibfield  {author} {\bibinfo {author} {\bibfnamefont {C.}~\bibnamefont
  {Elouard}}, \bibinfo {author} {\bibfnamefont {N.}~\bibnamefont {Bernardes}},
  \bibinfo {author} {\bibfnamefont {A.}~\bibnamefont {Carvalho}}, \bibinfo
  {author} {\bibfnamefont {M.}~\bibnamefont {Santos}},\ and\ \bibinfo {author}
  {\bibfnamefont {A.}~\bibnamefont {Auff{\`e}ves}},\ }\bibfield  {title}
  {\bibinfo {title} {Probing quantum fluctuation theorems in engineered
  reservoirs},\ }\href@noop {} {\bibfield  {journal} {\bibinfo  {journal} {New
  Journal of Physics}\ }\textbf {\bibinfo {volume} {19}},\ \bibinfo {pages}
  {103011} (\bibinfo {year} {2017}{\natexlab{a}})}\BibitemShut {NoStop}%
\bibitem [{\citenamefont {Elouard}\ \emph
  {et~al.}(2017{\natexlab{b}})\citenamefont {Elouard}, \citenamefont
  {Herrera-Mart{\'\i}}, \citenamefont {Clusel},\ and\ \citenamefont
  {Auff{\`e}ves}}]{elouard2017role}%
  \BibitemOpen
  \bibfield  {author} {\bibinfo {author} {\bibfnamefont {C.}~\bibnamefont
  {Elouard}}, \bibinfo {author} {\bibfnamefont {D.~A.}\ \bibnamefont
  {Herrera-Mart{\'\i}}}, \bibinfo {author} {\bibfnamefont {M.}~\bibnamefont
  {Clusel}},\ and\ \bibinfo {author} {\bibfnamefont {A.}~\bibnamefont
  {Auff{\`e}ves}},\ }\bibfield  {title} {\bibinfo {title} {The role of quantum
  measurement in stochastic thermodynamics},\ }\href@noop {} {\bibfield
  {journal} {\bibinfo  {journal} {npj Quantum Information}\ }\textbf {\bibinfo
  {volume} {3}},\ \bibinfo {pages} {1} (\bibinfo {year}
  {2017}{\natexlab{b}})}\BibitemShut {NoStop}%
\bibitem [{\citenamefont {Dressel}\ \emph {et~al.}(2017)\citenamefont
  {Dressel}, \citenamefont {Chantasri}, \citenamefont {Jordan},\ and\
  \citenamefont {Korotkov}}]{dressel2017arrow}%
  \BibitemOpen
  \bibfield  {author} {\bibinfo {author} {\bibfnamefont {J.}~\bibnamefont
  {Dressel}}, \bibinfo {author} {\bibfnamefont {A.}~\bibnamefont {Chantasri}},
  \bibinfo {author} {\bibfnamefont {A.~N.}\ \bibnamefont {Jordan}},\ and\
  \bibinfo {author} {\bibfnamefont {A.~N.}\ \bibnamefont {Korotkov}},\
  }\bibfield  {title} {\bibinfo {title} {Arrow of time for continuous quantum
  measurement},\ }\href@noop {} {\bibfield  {journal} {\bibinfo  {journal}
  {Physical Review Letters}\ }\textbf {\bibinfo {volume} {119}},\ \bibinfo
  {pages} {220507} (\bibinfo {year} {2017})}\BibitemShut {NoStop}%
\bibitem [{\citenamefont {Monsel}\ \emph {et~al.}(2018)\citenamefont {Monsel},
  \citenamefont {Elouard},\ and\ \citenamefont
  {Auff{\`e}ves}}]{monsel2018autonomous}%
  \BibitemOpen
  \bibfield  {author} {\bibinfo {author} {\bibfnamefont {J.}~\bibnamefont
  {Monsel}}, \bibinfo {author} {\bibfnamefont {C.}~\bibnamefont {Elouard}},\
  and\ \bibinfo {author} {\bibfnamefont {A.}~\bibnamefont {Auff{\`e}ves}},\
  }\bibfield  {title} {\bibinfo {title} {An autonomous quantum machine to
  measure the thermodynamic arrow of time},\ }\href@noop {} {\bibfield
  {journal} {\bibinfo  {journal} {npj Quantum Information}\ }\textbf {\bibinfo
  {volume} {4}},\ \bibinfo {pages} {1} (\bibinfo {year} {2018})}\BibitemShut
  {NoStop}%
\bibitem [{\citenamefont {Manikandan}\ \emph {et~al.}(2019)\citenamefont
  {Manikandan}, \citenamefont {Elouard},\ and\ \citenamefont
  {Jordan}}]{manikandan2019fluctuation}%
  \BibitemOpen
  \bibfield  {author} {\bibinfo {author} {\bibfnamefont {S.~K.}\ \bibnamefont
  {Manikandan}}, \bibinfo {author} {\bibfnamefont {C.}~\bibnamefont
  {Elouard}},\ and\ \bibinfo {author} {\bibfnamefont {A.~N.}\ \bibnamefont
  {Jordan}},\ }\bibfield  {title} {\bibinfo {title} {Fluctuation theorems for
  continuous quantum measurements and absolute irreversibility},\ }\href@noop
  {} {\bibfield  {journal} {\bibinfo  {journal} {Physical Review A}\ }\textbf
  {\bibinfo {volume} {99}},\ \bibinfo {pages} {022117} (\bibinfo {year}
  {2019})}\BibitemShut {NoStop}%
\bibitem [{\citenamefont {Belenchia}\ \emph {et~al.}(2020)\citenamefont
  {Belenchia}, \citenamefont {Mancino}, \citenamefont {Landi},\ and\
  \citenamefont {Paternostro}}]{Belenchia20}%
  \BibitemOpen
  \bibfield  {author} {\bibinfo {author} {\bibfnamefont {A.}~\bibnamefont
  {Belenchia}}, \bibinfo {author} {\bibfnamefont {L.}~\bibnamefont {Mancino}},
  \bibinfo {author} {\bibfnamefont {G.}~\bibnamefont {Landi}},\ and\ \bibinfo
  {author} {\bibfnamefont {M.}~\bibnamefont {Paternostro}},\ }\bibfield
  {title} {\bibinfo {title} {Entropy production in continuously measured
  gaussian quantum systems},\ }\href@noop {} {\bibfield  {journal} {\bibinfo
  {journal} {npj {Quantum} {Information}}\ }\textbf {\bibinfo {volume} {97}}
  (\bibinfo {year} {2020})}\BibitemShut {NoStop}%
\bibitem [{\citenamefont {Matos}\ \emph {et~al.}(2022)\citenamefont {Matos},
  \citenamefont {Kantorovich},\ and\ \citenamefont {Ford}}]{matos2022}%
  \BibitemOpen
  \bibfield  {author} {\bibinfo {author} {\bibfnamefont {D.}~\bibnamefont
  {Matos}}, \bibinfo {author} {\bibfnamefont {L.}~\bibnamefont {Kantorovich}},\
  and\ \bibinfo {author} {\bibfnamefont {I.~J.}\ \bibnamefont {Ford}},\
  }\bibfield  {title} {\bibinfo {title} {Stochastic entropy production for
  continuous measurements of an open quantum system},\ }\href@noop {}
  {\bibfield  {journal} {\bibinfo  {journal} {Journal of Physics
  Communications}\ }\textbf {\bibinfo {volume} {6}},\ \bibinfo {pages} {125003}
  (\bibinfo {year} {2022})}\BibitemShut {NoStop}%
\bibitem [{\citenamefont {Gardiner}(2009)}]{gardiner2009stochastic}%
  \BibitemOpen
  \bibfield  {author} {\bibinfo {author} {\bibfnamefont {C.}~\bibnamefont
  {Gardiner}},\ }\href@noop {} {\emph {\bibinfo {title} {Handbook of Stochastic
  Methods}}},\ Vol.~\bibinfo {volume} {4}\ (\bibinfo  {publisher} {Springer
  Berlin},\ \bibinfo {year} {2009})\BibitemShut {NoStop}%
\bibitem [{\citenamefont {{Schmidt}}\ \emph {et~al.}(2015)\citenamefont
  {{Schmidt}}, \citenamefont {{Carusela}}, \citenamefont {{Pekola}},
  \citenamefont {{Suomela}},\ and\ \citenamefont {{Ankerhold}}}]{Schmidt}%
  \BibitemOpen
  \bibfield  {author} {\bibinfo {author} {\bibfnamefont {R.}~\bibnamefont
  {{Schmidt}}}, \bibinfo {author} {\bibfnamefont {M.~F.}\ \bibnamefont
  {{Carusela}}}, \bibinfo {author} {\bibfnamefont {J.~P.}\ \bibnamefont
  {{Pekola}}}, \bibinfo {author} {\bibfnamefont {S.}~\bibnamefont
  {{Suomela}}},\ and\ \bibinfo {author} {\bibfnamefont {J.}~\bibnamefont
  {{Ankerhold}}},\ }\bibfield  {title} {\bibinfo {title} {{Work and heat for
  two-level systems in dissipative environments: Strong driving and
  non-Markovian dynamics}},\ }\href
  {https://doi.org/10.1103/PhysRevB.91.224303} {\bibfield  {journal} {\bibinfo
  {journal} {Physical Review B}\ }\textbf {\bibinfo {volume} {91}},\ \bibinfo
  {eid} {224303} (\bibinfo {year} {2015})}\BibitemShut {NoStop}%
\bibitem [{\citenamefont {Lindblad}(1976)}]{lindblad}%
  \BibitemOpen
  \bibfield  {author} {\bibinfo {author} {\bibfnamefont {G.}~\bibnamefont
  {Lindblad}},\ }\bibfield  {title} {\bibinfo {title} {On the generators of
  quantum dynamical semigroups},\ }\href@noop {} {\bibfield  {journal}
  {\bibinfo  {journal} {Communications in Mathematical Physics}\ }\textbf
  {\bibinfo {volume} {48}},\ \bibinfo {pages} {119} (\bibinfo {year}
  {1976})}\BibitemShut {NoStop}%
\bibitem [{\citenamefont {Moodley}\ and\ \citenamefont
  {Petruccione}(2009)}]{moodley2009stochastic}%
  \BibitemOpen
  \bibfield  {author} {\bibinfo {author} {\bibfnamefont {M.}~\bibnamefont
  {Moodley}}\ and\ \bibinfo {author} {\bibfnamefont {F.}~\bibnamefont
  {Petruccione}},\ }\bibfield  {title} {\bibinfo {title} {{Stochastic
  wave-function unraveling of the generalized Lindblad master equation}},\
  }\href@noop {} {\bibfield  {journal} {\bibinfo  {journal} {Physical Review
  A}\ }\textbf {\bibinfo {volume} {79}},\ \bibinfo {pages} {042103} (\bibinfo
  {year} {2009})}\BibitemShut {NoStop}%
\bibitem [{\citenamefont {Yan}\ and\ \citenamefont
  {Shao}(2016)}]{yan2016stochastic}%
  \BibitemOpen
  \bibfield  {author} {\bibinfo {author} {\bibfnamefont {Y.-A.}\ \bibnamefont
  {Yan}}\ and\ \bibinfo {author} {\bibfnamefont {J.}~\bibnamefont {Shao}},\
  }\bibfield  {title} {\bibinfo {title} {Stochastic description of quantum
  {Brownian} dynamics},\ }\href@noop {} {\bibfield  {journal} {\bibinfo
  {journal} {Frontiers of Physics}\ }\textbf {\bibinfo {volume} {11}},\
  \bibinfo {pages} {1} (\bibinfo {year} {2016})}\BibitemShut {NoStop}%
\bibitem [{\citenamefont {Ford}(2015)}]{Ford}%
  \BibitemOpen
  \bibfield  {author} {\bibinfo {author} {\bibfnamefont {I.~J.}\ \bibnamefont
  {Ford}},\ }\bibfield  {title} {\bibinfo {title} {Measures of thermodynamic
  irreversibility in deterministic and stochastic dynamics},\ }\href
  {http://stacks.iop.org/1367-2630/17/i=7/a=075017} {\bibfield  {journal}
  {\bibinfo  {journal} {New Journal of Physics}\ }\textbf {\bibinfo {volume}
  {17}},\ \bibinfo {pages} {075017} (\bibinfo {year} {2015})}\BibitemShut
  {NoStop}%
\bibitem [{\citenamefont {Crooks}(2008)}]{crooks2008quantum}%
  \BibitemOpen
  \bibfield  {author} {\bibinfo {author} {\bibfnamefont {G.~E.}\ \bibnamefont
  {Crooks}},\ }\bibfield  {title} {\bibinfo {title} {Quantum operation time
  reversal},\ }\href@noop {} {\bibfield  {journal} {\bibinfo  {journal}
  {Physical Review A}\ }\textbf {\bibinfo {volume} {77}},\ \bibinfo {pages}
  {034101} (\bibinfo {year} {2008})}\BibitemShut {NoStop}%
\bibitem [{\citenamefont {Ford}(2016)}]{Ford16}%
  \BibitemOpen
  \bibfield  {author} {\bibinfo {author} {\bibfnamefont {I.~J.}\ \bibnamefont
  {Ford}},\ }\bibfield  {title} {\bibinfo {title} {Maxwell's demon and the
  management of ignorance in stochastic thermodynamics},\ }\href@noop {}
  {\bibfield  {journal} {\bibinfo  {journal} {Contemporary Physics}\ }\textbf
  {\bibinfo {volume} {57}},\ \bibinfo {pages} {309} (\bibinfo {year}
  {2016})}\BibitemShut {NoStop}%
\end{thebibliography}%

\end{document}